\documentclass[seceq]{ptptex}

\usepackage{graphicx}
\renewcommand{\vec}[1]{\mbox{\boldmath $#1$}}

\title{
Subbarrier fusion reactions and many-particle quantum tunneling
}

\author{Kouichi \textsc{Hagino}$^1$ and Noboru \textsc{Takigawa}$^{1,2}$}
\inst{$^1$Department of Physics, Tohoku University, Sendai 980-8578, Japan \\
$^2$Tohoku Instiute of Technology, Sendai 982-8577, Japan}

\abst{%         %This abstract is neglected when [addenda] or [errata].
Low energy heavy-ion fusion reactions 
are governed by quantum tunneling through 
the Coulomb barrier 
formed by a strong cancellation 
of the repulsive Coulomb force with the attractive nuclear 
interaction between the colliding nuclei. 
Extensive experimental as well as theoretical studies have revealed 
that fusion reactions are strongly influenced by 
couplings of the relative motion of the 
colliding nuclei to several nuclear intrinsic motions. 
Heavy-ion subbarrier fusion reactions thus provide a good 
opportunity to address a general problem on quantum tunneling in the 
presence of couplings, which has been a popular subject in 
the past decades in many branches of physics and chemistry. 
Here we review theoretical aspects of heavy-ion subbarrier fusion 
reactions from the view point of quantum tunneling in systems 
with many degrees 
of freedom. 
Particular emphases are put on the coupled-channels approach to 
fusion reactions, and the barrier distribution representation for 
multi-channel penetrability. We also 
discuss an application of
the barrier distribution
method to elucidation of the mechanism of dissociative 
adsorption of H$_2$ molecules in surface
science. 
}

\begin{document}

\maketitle

\section{Introduction}

Quantum mechanics is indispensable in understanding 
microscopic systems such as atoms, molecules, 
and atomic nuclei. 
One of its fundamental aspects is a quantum tunneling, where a 
particle penetrates into a classically forbidden region. 
This is a wave phenomenon and is frequently 
encountered in 
diverse processes in physics and chemistry. 

The importance of quantum tunneling has been recognized from 
the birth of quantum mechanics. 
For instance, it was as early as 1928 when 
Gamow, and independently Gurney and Condon, applied quantum tunneling 
to $\alpha$ decays of atomic nuclei and 
successfully explained 
the systematics of the experimental half-lives of radioactive 
nuclei\cite{G28,GC28}. 

In many applications of quantum tunneling, one only 
considers penetration of a one-dimensional 
potential barrier, or a barrier with a single variable. 
In general, however, a particle which penetrates a potential barrier  
is never isolated but interacts with 
its surroundings or environments, resulting in modification in 
its behavior. Moreover, when the particle is a composite particle, 
the quantum tunneling has to be discussed from a many-particle 
point of view. 
Quantum tunneling therefore inevitably takes 
place in reality in a multi-dimensional space. 
Such problem was first addressed 
by Kapur and Peierls in 1937 \cite{KP37}. Their theory has further 
been developed by {\it e.g.,} 
Banks, Bender, and Wu \cite{BBW73}, Gervais and 
Sakita \cite{GS77}, Brink, Nemes, and Vautherin \cite{BNV83}, 
Schmid \cite{S86}, and Takada and Nakamura \cite{TN94}. 

When the quantum tunneling occurs in a complex system, 
such as the trapped flux in a 
superconducting quantum interference devices (SQUID) ring \cite{SSAL85}, 
the tunneling variable couples to a large number of other degrees of freedom. 
In such systems, the environmental 
degrees of freedom more or less reveal a dissipative character. 
Quantum tunneling under the influence of dissipative environments
plays an important role and is a fundamental problem in many fields 
of physics and chemistry. 
This problem has been studied in detail by Caldeira and Leggett \cite{CL81}. 
This seminal work has stimulated lots of 
experimental and theoretical works, and has made 
quantum tunneling in systems with many degrees of freedom 
a topic of immense interest during the past decades \cite{JJAP93}. 

\begin{figure}[t]
\centerline{\includegraphics[width=7cm,clip]{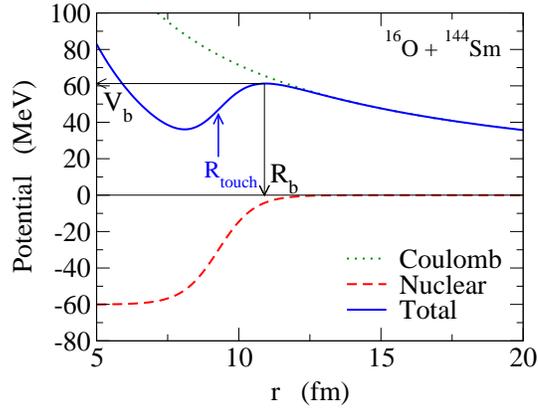}}
\caption{
The internucleus potential between 
$^{16}$O and $^{144}$Sm nuclei 
as a function of the relative distance. 
The dotted and the dashed lines are the Coulomb and the nuclear potentials, 
respectively, while the solid line denotes the total potential. 
$V_b$ and $R_b$ are the height and the position of 
the Coulomb barrier, respectively. $R_{\rm touch}$ is 
the touching radius at which 
the projectile and the target nuclei start overlapping significantly 
with each other. }
\end{figure}

In nuclear physics, 
one of the typical examples of tunneling phenomena 
is heavy-ion 
fusion reaction at energies near and below 
the Coulomb barrier\cite{BT98,DHRS98}. 
Fusion is defined as a reaction in which two separate nuclei combine together 
to form a compound nucleus. 
In order for fusion reaction to take place, the relative motion 
between the colliding nuclei has to overcome the Coulomb barrier 
formed by a strong cancellation between the long-ranged repulsive 
Coulomb and the short-ranged attractive nuclear forces 
(as a typical example, Fig.1 shows the internucleus potential 
between $^{16}$O and $^{144}$Sm nuclei 
as a function of the relative distance). 
Unless under extreme conditions, 
it is reasonable to assume that atomic nuclei are 
isolated systems and the couplings to external environments 
can be neglected. 
Nevertheless, one can still consider 
{\it intrinsic} environments. 
The whole spectra of excited states 
of the target and projectile nuclei (as well as several sorts of 
nucleon transfer processes) are populated in a complex way during  
fusion reactions. They act as environments 
to which 
the relative motion between the colliding nuclei 
couples.  
In fact, it has by now been well established that 
cross sections of heavy-ion fusion reactions are substantially 
enhanced due to couplings to nuclear intrinsic degrees of freedom 
at energies below the Coulomb barrier 
as compared to the predictions of a simple potential model 
\cite{BT98,DHRS98,B88,SRB86,R94,S78,LDH95}.  
Heavy-ion subbarrier fusion reactions thus make good examples 
of environment-assisted tunneling phenomena. 

Theoretically the standard way to address the effects of the 
couplings between the relative motion and nuclear intrinsic 
degrees of freedom on fusion reactions 
is to numerically solve the 
coupled-channels equations which include all the relevant channels. 
In the eigen-channel representation of coupled-channels equations, 
the channel coupling effects can be interpreted in terms of 
a distribution of fusion 
barriers\cite{DHRS98,DLW83,E81,NBT86}. 
In this representation, the 
fusion cross section is given 
by a weighted sum of the fusion cross sections for each eigen-barrier. 
Those eigen-barriers lower than the original barrier 
are responsible for the enhancement of the 
fusion cross section at energies below the Coulomb barrier. 
Based on this idea, Rowley, Satchler, and 
Stelson have proposed a method to extract barrier distributions directly from 
experimental fusion excitation functions 
by taking the second derivative of the product of the fusion cross 
section and the center of mass energy $E \sigma_{\rm fus}$ with respect to
$E$, {\it i.e.,} $d^2(E\sigma_{\rm fus})/dE^2$ ~\cite{RSS91}. 
This method was tested against high precision 
experimental data of fusion cross sections soon after the 
method was proposed\cite{WLH91}. 
The extracted fusion barrier distributions 
were sensitive to the effects of channel-couplings and 
provided a much more apparent way of understanding their effects on the 
fusion process than the fusion excitation functions themselves. 
It is now well recognised that 
the barrier distribution approach is 
a standard tool for heavy-ion subbarrier fusion reactions \cite{DHRS98,LDH95}. 

The aim of this paper is to review 
theoretical aspects of heavy-ion subbarrier fusion reactions 
from the view point of quantum tunneling of composite particles. 
To this end, we mainly base our discussions 
on the coupled-channels approach.  Earlier reviews on the subbarrier 
fusion reactions can be found in Refs. 
\citen{BT98,DHRS98,B88,SRB86,R94}. 
See also Refs. \citen{CGDH06} and \citen{LS05} for reviews on subbarrier fusion 
reactions of radioactive nuclei, and {\it e.g.,} Refs. \citen{A00} and 
\citen{AM12} 
for reviews on fusion reactions relevant to synthesis of superheavy 
elements, both of which we do not cover in this article. 

The paper is organized as follows.  
We will first discuss in the next section a potential model approach to 
heavy-ion fusion reactions. This is the simplest approach to fusion 
reaction, in which only elastic scattering and fusion are assumed to 
occur. This approach is adequate for light systems, but for 
fusion with a medium-heavy or heavy target nucleus the effects of 
nuclear excitations during fusion start playing an important role. 
In Sec. 3, we will 
discuss such nuclear structure effect on heavy-ion fusion reactions. 
To this end, we will introduce and detail 
the coupled-channels formalism 
which takes into account the inelastic scattering 
and transfer processes during fusion reactions. 
In Sec. 4, 
light will be shed on the fusion barrier distribution representation 
of fusion cross section defined as $d^2(E\sigma_{\rm fus})/dE^2$. 
It has been known that this approach is exact when 
the excitation energy of the intrinsic motion is zero, but we will 
demonstrate that one can generalize it unambiguously using 
the eigen-channel approach also to the case when the excitation 
energy is finite. 
In Sec. 5, we will turn to a discussion on the present 
status of our understanding 
of deep subbarrier fusion reactions. 
At these energies, fusion cross sections have been shown 
to be suppressed compared 
to the values of the 
standard coupled-channels calculations. This phenomenon 
may be related to dissipative quantum tunneling, that is, an irreversible 
coupling to intrinsic degrees of freedom. 
In Sec. 6, we discuss an application of 
the barrier distribution method to surface physics, more specifically, 
the effect of rotational excitations on 
a dissociative adsorption process of H$_2$ molecules. We then 
summarize the 
paper in Sec. 7. 

\section{One dimensional potential model}

\subsection{Ion-ion potential}

Theoretically, the simplest approach to heavy-ion fusion reactions 
is to use the one dimensional potential model where both the projectile 
and the target are assumed to be structureless. 
A potential between the projectile and the target is given 
by a function of the relative distance $r$ between them. It consists 
of two parts, that is, 
\begin{equation}
V(r)=V_N(r) + V_C(r), 
\end{equation}
where $V_N(r)$ is the nuclear potential, and $V_C(r)$ is the Coulomb 
potential given by 
\begin{equation}
V_C(r)=\frac{Z_PZ_Te^2}{r},
\end{equation}
in the outside region where the projectile and the target nuclei do not 
significantly overlap with each other. 
Figure 1 shows a typical potential $V(r)$ for the $s$-wave scattering 
of the $^{16}$O + $^{144}$Sm reaction. The dotted and the dashed lines are 
the nuclear and the Coulomb potentials, respectively, while the total 
potential $V(r)$ is denoted by the solid line. 
One can see that a potential barrier appears due to a strong cancellation 
between the short-ranged attractive nuclear interaction and the 
long-ranged repulsive Coulomb force. This potential barrier is referred 
to as the 
{\it Coulomb barrier} and has to be overcome in order for the fusion reaction 
to take place. 
$R_{\rm touch}$ in the figure is the touching radius, at which the 
projectile and the target nuclei begin overlapping considerably. 
One can see that the 
Coulomb barrier is located outside the touching radius. 

There are several ways to estimate the nuclear potential $V_N(r)$. 
One standard method is to fold a nucleon-nucleon interaction with 
the projectile 
and the target densities \cite{SL79}. 
The direct part of the 
nuclear potential in this double folding procedure is given by 
\begin{equation}
V_N(r)=\int d \vec{r}_1 d \vec{r}_2\,
v_{NN}(\vec{r}_2 - \vec{r}_1 - 
\vec{r}) \rho_P(\vec{r}_1)
\rho_T(\vec{r}_2),
\end{equation}
where $v_{NN}$ is an effective nucleon-nucleon interaction, 
and $\rho_P$ and $\rho_T$ 
are the densities of the projectile and the target, respectively. 
The double-folding potential is in general a non-local potential 
due to the anti-symmetrization effect of nucleons. 
Usually, either a zero-range approximation \cite{SL79,BS97} or 
a local momentum approximation \cite{KS00,KSO97,K01,S75,SM79} 
is employed in order to treat 
the non-locality of the potential. 

A phenomenological nuclear potential has also been employed. 
For instance, 
a Woods-Saxon form 
\begin{equation}
V_N(r)=-\frac{V_0}{1+\exp[(r-R_0)/a]},
\label{WS}
\end{equation}
with 
\begin{eqnarray}
V_0&=&16\pi\gamma\bar{R}a, \\
R_0&=&R_P + R_T, \\
R_i &=& 1.20 A_i^{1/3} - 0.09~{\rm fm}~~~~~~(i= P, T), \\
\bar{R} &=& R_PR_T/(R_P + R_T), \\
\gamma &=& 0.95 
\left[1-1.8 \left(\frac{N_P - Z_P}{A_P}\right)
\left(\frac{N_T - Z_T}{A_T}\right)\right]~{\rm MeV~fm^{-2}},\\
1/a&=&1.17\left[1+0.53\left(A_P^{-1/3}+A_T^{-1/3}\right)\right]~{\rm fm^{-1}}, 
\end{eqnarray}
has been widely used, where the parameters were determined from 
a least-squares fit to the experimental data of heavy-ion 
elastic scattering \cite{BW91,AW79}. 

A nuclear potential so constructed 
has been successful in reproducing experimental 
angular distributions of elastic and inelastic scattering 
for many systems. Moreover, the empirical value of surface diffuseness 
parameter, $a\sim$ 0.63 fm, is consistent with a double folding 
potential. 
Recently, a value of the surface diffuseness parameter has been determined 
unambiguously using heavy-ion quasi-elastic scattering at deep subbarrier 
energies\cite{HTBT05,WKD06}. It has been confirmed that 
the experimental data are consistent with a value around 
$a\sim$ 0.63 fm \cite{WKD06,GEH07,EDH08,LJZ09}. 

In marked contrast, recent experimental data
for heavy-ion subbarrier fusion reactions 
suggest that a much larger value of diffuseness, ranging
from 0.75 to 1.5 fm, is required to fit the data 
\cite{LDH95,NMD01,NBD04,MHD07,GHDN04,HRD03}. 
The Woods-Saxon 
potential which fits elastic scattering overestimates fusion
cross sections at energies both above and below the Coulomb
barrier, having an inconsistent energy dependence with the
experimental fusion excitation function. 
A reason for the large discrepancies in diffuseness
parameters extracted from scattering and fusion analyses 
has not yet been fully understood.
However, it is probably the case that the double folding procedure 
is valid only in the surface region, while  
several dynamical effects come into play in the inner part where 
fusion is sensitive to. 

We summarize the relation between the surface diffuseness parameter $a$ 
of a nuclear potential 
and the parameters of the Coulomb barrier, that is, the curvature, 
the barrier height, and the barrier position in Appendix A for 
an exponential and a Woods-Saxon potentials. 

\subsection{Fusion cross sections}

In the potential model, the internucleus potential, $V(r)$, is supplemented 
by an imaginary part, $-iW(r)$, which mocks up the formation of a compound 
nucleus. One then solves the Schr\"odinger equation 
\begin{equation}
\left[-\frac{\hbar^2}{2\mu}\frac{d^2}{dr^2}
+V(r)-iW(r)+\frac{l(l+1)\hbar^2}{2\mu r^2}
-E\right]
u_l(r)=0,
\end{equation}
for each partial wave $l$, 
where $\mu$ is the reduced mass of the system, with the boundary 
conditions of 
\begin{eqnarray}
u_l(r)&\sim& r^{l+1} ~~~~~~~~~~~~~~~~~~~~~~~~~~~~~~~r\to 0, 
\label{boundary}
\\
&=&H_l^{(-)}(kr)-S_l \,H_l^{(+)}(kr) ~~~~~~r\to \infty.
\label{boundary2}
\end{eqnarray}
Here, $H_l^{(+)}$ and $H_l^{(-)}$ 
are the outgoing and the incoming Coulomb wave functions, respectively. 
$S_l$ is the nuclear $S$-matrix, and $k=\sqrt{2\mu E/\hbar^2}$ is the 
wave number associated with the energy $E$. 

If the imaginary part of the potential, $W(r)$, is confined well inside 
the Coulomb barrier, one can regard the total absorption cross section as 
fusion cross section, {\it i.e.,} 
\begin{equation}
\sigma_{\rm fus}(E)\sim\sigma_{\rm abs}(E)=
\frac{\pi}{k^2}
\sum_l (2l+1)\left(1-|S_l|^2\right).
\label{fuscross}
\end{equation}

In heavy-ion fusion reactions, instead of imposing the regular boundary 
condition at the origin, Eq. (\ref{boundary}), 
the so called incoming wave boundary 
condition (IWBC) is often applied 
without introducing the imaginary part of the potential, $W(r)$ 
\cite{DLW83,LP84}. 
Under the incoming wave boundary condition, 
the wave function has a form 
\begin{equation}
u_l(r)=\sqrt{\frac{k}{k_l(r)}}\,{\cal T}_l\exp\left(-i\int^r_{r_{\rm abs}}k_l(r')dr'\right) 
~~~~~r \leq r_{\rm abs},
\label{iwbc}
\end{equation}
at the distance smaller than the absorption radius $r_{\rm abs}$, which is taken 
to be inside the Coulomb barrier. Here, $k_l(r)$ is the local wave 
number for the $l$-th partial wave defined by 
\begin{equation}
k_l(r)=\sqrt{\frac{2\mu}{\hbar^2}\left(E-V(r)
-\frac{l(l+1)\hbar^2}{2\mu r^2}\right)}.
\end{equation}
The incoming wave boundary condition corresponds 
to the case where there is a strong absorption in the inner region so that 
the incoming flux never returns back. 
For heavy-ion fusion reactions, the final result is not sensitive to 
the choice of the absorption radius $r_{\rm abs}$, and it is often taken to 
be at the pocket of the potential\cite{HRK99}. 
With the incoming wave boundary condition, ${\cal T}_l$ in Eq. (\ref{iwbc}) is 
interpreted as the transmission coefficient. 
Equation (\ref{fuscross}) 
is then transformed to 
\begin{equation}
\sigma_{\rm fus}(E)=\frac{\pi}{k^2}\sum_l(2l+1)P_l(E),
\label{fuscross2}
\end{equation}
where $P_l(E)$ is the penetrability for the $l$-wave scattering 
defined as 
\begin{equation}
P_l(E)=1-|S_l|^2=\left|{\cal T}_l\right|^2,
\end{equation}
for the boundary conditions (\ref{boundary2}) and (\ref{iwbc}). 
The mean angular momentum of the compound nucleus is 
evaluated in a similar way as 
\begin{equation}
\langle l\rangle (E) = \frac{\frac{\pi}{k^2}\sum_ll(2l+1)P_l(E)}
{\frac{\pi}{k^2}\sum_l(2l+1)P_l(E)}.
\end{equation}
For a parabolic potential, Wong has derived an analytic expression 
for fusion cross sections, Eq. (\ref{fuscross2})\cite{W73}. 
We will discuss it in Appendix B. 

The incoming wave boundary condition, Eq. (\ref{iwbc}), 
has two advantages over the regular boundary 
condition, Eq. (\ref{boundary}). The first advantage is that 
the imaginary part of nuclear potential is not needed, and 
the number of adjustable parameters can be reduced. 
The second point is that the incoming wave boundary condition directly 
provides the penetrability $P_l(E)=\left|{\cal T}_l\right|^2$ 
and thus the round off error 
can be avoided in evaluating $1-|S_l|^2$. This is a crucial point at 
energies well below the Coulomb barrier, where $S_l$ is close to unity. 
Notice that the incoming wave boundary condition does not 
necessarily correspond 
to the limit of $W(r)\to \infty$, as the quantum reflection due to $W(r)$ 
has to be neglected in order to realize it. The incoming wave boundary 
condition should thus be regarded as a different model from the regular 
boundary condition. 

\subsection{Comparison with experimental data: success and failure of 
the potential model}

\begin{figure}[t]
\centerline{\includegraphics[width=11cm,clip]{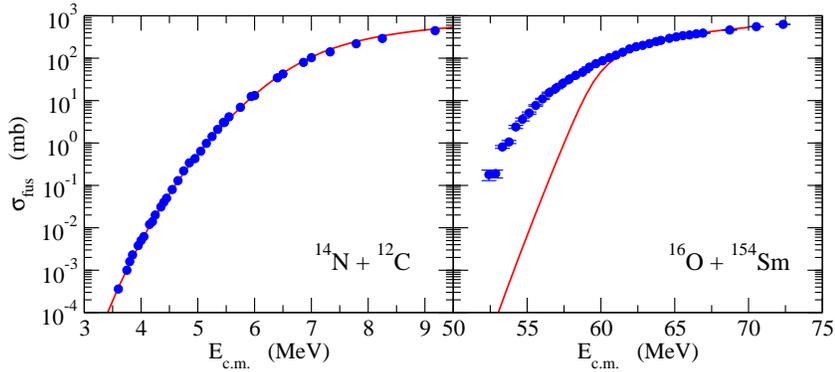}}
\caption{
Comparison of experimental fusion cross sections for the 
$^{14}$N+$^{12}$C system (the left panel) and 
$^{16}$O+$^{154}$Sm system (the right panel) with results of the 
potential model calculations (the solid line). 
The height of the Coulomb barrier is around $V_b\sim$ 
6.9 MeV and 59 MeV for 
$^{14}$N+$^{12}$C 
and $^{16}$O+$^{154}$Sm, respectively. 
The experimental data are taken from Refs. \citen{SSW77} and \citen{LDH95} 
for the $^{14}$N+$^{12}$C and the $^{16}$O+$^{154}$Sm reactions, respectively. 
}
\end{figure}

Let us now compare the one dimensional potential model for heavy-ion 
fusion reaction with experimental data. 
Figure 2 shows 
the experimental excitation functions of fusion 
cross section for $^{14}$N+$^{12}$C (the left panel) and 
$^{16}$O+$^{154}$Sm (the right panel) systems, as well as  
results of the potential model calculation (the solid lines). 
One can see that the potential model well reproduces 
the experimental data for the lighter system, 
$^{14}$N + $^{12}$C. 
On the other hand, 
the potential model apparently underestimates 
fusion cross sections for the heavier system, 
$^{16}$O + $^{154}$Sm, although it reproduces the experimental data 
at energies above the Coulomb barrier, which is about 
59 MeV for this system. 

\begin{figure}[bt]
\centerline{\includegraphics[width=7cm,clip]{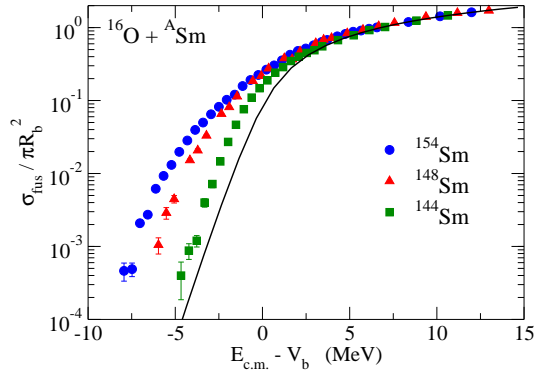}}
\caption{
The experimental fusion cross sections for 
$^{16}$O+$^{144,148,154}$Sm systems, taken from Ref. \citen{LDH95}. 
In order to remove the trivial target dependence, the experimental 
fusion cross sections are 
divided by $\pi R_b^2$, where $R_b$ is the position of the Coulomb barrier, 
and the energies are measured with respect to the barrier height, $V_b$, 
for each system. The solid line shows the result of the potential 
model calculation.}
\end{figure}

In order to understand the origin for the failure of the potential model, 
Fig. 3 shows the experimental fusion 
excitation functions for $^{16}$O + $^{144,148,154}$Sm 
reactions \cite{LDH95} and a comparison with the potential model 
(the solid line). 
To remove trivial target dependence, 
these are plotted as a function of center of 
mass energy relative to the barrier height for each system, 
and the fusion cross sections are divided by the geometrical factor, 
$\pi R_b^2$. With these prescriptions, the fusion cross sections 
for the different systems 
are matched with each other at energies above the Coulomb barrier, 
although one can also consider a more refined 
prescription\cite{CGLCC09,GLC09}.   
The barrier height and the result of the potential model are obtained 
with the Aky\"uz-Winther potential\cite{AW79}. 
One again observes that the experimental fusion cross sections are drastically 
enhanced at energies below the Coulomb barrier 
compared with the prediction of the potential model. 
Moreover, one also observes that the degree of enhancement of fusion 
cross section depends strongly on the target nucleus. That is, 
the enhancement for 
the $^{16}$O + $^{154}$Sm system is order of magnitude, while that 
for the $^{16}$O + $^{144}$Sm system is about a factor of four at 
energies below the Coulomb barrier. 
This strong target dependence of fusion cross sections suggests that 
low-lying collective excitations play a role, as we will discuss in the next 
section. 

The inadequacy of the potential model has been demonstrated in a more 
transparent way by Balantekin {\it et al.}\cite{BKN83}. 
Within the semi-classical approximation, 
the penetrability for a one-dimensional 
barrier can be inverted to yield the barrier thickness \cite{CG78}. 
Balantekin {\it et al.} applied such inversion formula directly to 
the experimental fusion cross sections 
in order to construct an effective internucleus potential. 
Assuming a one-dimensional 
energy-independent local potential, the resultant potentials 
were unphysically thin for heavy systems, often multi-valued potential. 
This result was confirmed also by 
the systematic study in Ref. \citen{IK84}. 
These analyses have provided a clear evidence
for the inadequacy of the one-dimensional barrier passing model
for heavy-ion fusion reactions, and has triggered to develop
the coupled-channels approach, which we will discuss in the next section.

In passing, 
we have recently applied 
the inversion procedure in a modified way 
to determine the lowest potential barrier 
among the distributed ones due to 
the effects of channel coupling \cite{HW07}. 
The extracted potential
for the $^{16}$O + $^{208}$Pb scattering 
is well behaved, indicating that the channel coupling 
indeed plays an essential role in subbarrier fusion reactions. 

\section{Coupled-channels formalism for heavy-ion fusion reactions}

\subsection{Nuclear structure effects on subbarrier fusion reactions}

\begin{figure}[t]
\centerline{\includegraphics[width=9cm,clip]{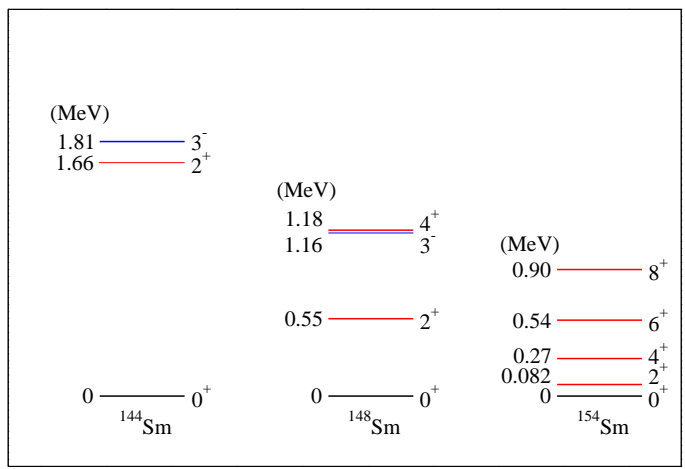}}
\caption{
The experimental low-lying spectra of $^{144,148,154}$Sm nuclei.}
\end{figure}

The strong target dependence of subbarrier fusion cross sections shown 
in Fig. 3 suggests that the enhancement of fusion cross sections is 
due to low-lying collective excitations of the colliding 
nuclei during fusion. The low-lying excited 
states in even-even nuclei are collective states, and strongly 
reflect the pairing correlation and shell structure. They have thus 
strongly coupled to the ground state, and also have a strong mass number 
and atomic number dependences. 
As an example, 
the low-lying spectra are shown in Fig. 4 for 
$^{144,148,154}$Sm. 
The $^{144}$Sm nucleus is close to the (sub-)shell closures ($Z$=64 and $N=82$) 
and is characterized by a strong octupole vibration. 
$^{154}$Sm, on the other hand, is a well deformed nucleus, and has a 
well developed ground state rotational band. 
$^{148}$Sm is a transitional nucleus, and there exists 
a soft quadrupole vibration 
in the low-lying spectrum. 
One can clearly see that there is a strong correlation 
between the degree of 
enhancement of fusion cross sections shown in Fig. 3 and 
{\it e.g.,} the energy of the first 2$^+$ state. 

Besides the low-lying collective excitations, there are many other 
modes of excitations in atomic nuclei. 
Among them, 
non-collective excitations 
couple only weakly to the ground state and usually they do not 
affect in a significant way 
heavy-ion fusion reactions, even though 
the number of non-collective states is large\cite{YHR10}.  
Couplings to giant resonances 
are relatively strong due to their collective character. 
However, since their excitation energies 
are relatively large and also are smooth functions 
of mass number \cite{BM75,BB94,HW01}, 
their effects can be effectively incorporated in 
a choice of internuclear potential through the adiabatic potential 
normalization (see the next section). 

\begin{figure}[t]
\centerline{\includegraphics[width=11cm,clip]{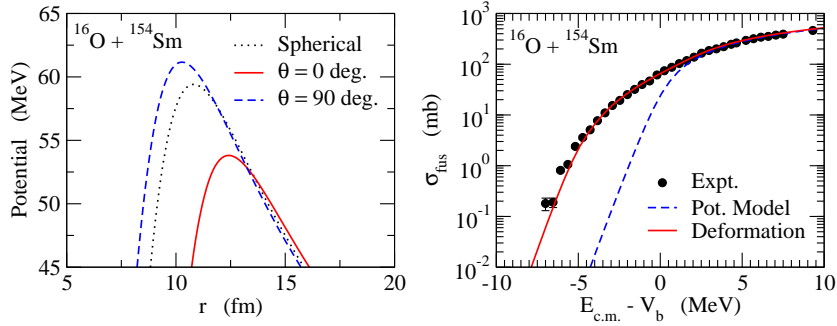}}
\caption{
(The left panel) the orientation dependence of fusion potential 
for the $^{16}$O+$^{154}$Sm reaction. The solid and the dashed lines 
are the potentials when the orientation of the deformed $^{154}$Sm target 
is $\theta=$ 0 and $\pi/2$, respectively. The dotted line denotes 
the potential when the deformation of $^{154}$Sm is not taken into account. 
(The right panel) the 
fusion cross sections for the $^{16}$O+$^{154}$Sm reaction. 
The dashed line is the result of the potential model calculation shown 
in Fig. 3, while the solid line is obtained by taking into account 
the deformation of the $^{154}$Sm nucleus with Eq. (\ref{orientation}). 
The experimental data are taken from Ref. \citen{LDH95}. 
}
\end{figure}

The effect of rotational excitations of a heavy 
deformed nucleus can be easily taken into account using the 
orientation average formula\cite{NBT86,W73,S78,SMBG75,RHT01,HR04}. 
For an axially symmetric target nucleus, fusion cross sections are 
computed with this formula as,
\begin{equation}
\sigma_{\rm fus}(E)=\int^1_0d(\cos\theta)\,\sigma_{\rm fus}(E;\theta),
\label{orientation}
\end{equation}
where $\theta$ is the angle between the symmetry axis and 
the beam direction. 
$\sigma_{\rm fus}(E;\theta)$ is a fusion cross section for a {\it fixed} 
orientation 
angle, $\theta$. This is obtained with {\it e.g.,} a deformed 
Woods-Saxon potential, 
\begin{equation}
V_N(r,\theta)=-\frac{V_0}{1+\exp[(r-R_0-R_T\beta_2Y_{20}(\theta)
-R_T\beta_4Y_{40}(\theta))/a]},
\label{defWS}
\end{equation}
which can be constructed by changing the target radius $R_T$ in the 
Woods-Saxon potential, Eq. (\ref{WS}), to 
$R_T\to R_T(1+\beta_2Y_{20}(\theta)+\beta_4Y_{40}(\theta))$. 
See Ref. \citen{AHNC12} for a recent application of this formula to 
fusion of massive systems, in which the formula is combined with classical 
Langevin calculations. 

The left panel of Fig. 5 shows the potential for the 
$^{16}$O+$^{154}$Sm reaction obtained with the deformation parameters 
of $\beta_2$=0.306 and $\beta_4=0.05$. 
The deformation of the Coulomb potential is also taken into account (see 
Sec. 3.4 for details). 
The solid line shows the potential for $\theta=0$. 
For this orientation angle, 
the potential is lowered 
by the deformation effect as compared to the spherical potential shown by the 
dotted line, because the attractive nuclear interaction is active from 
relatively large values of $r$. The opposite happens when $\theta =\pi/2$ 
as shown by the dashed line. The potential is distributed between the 
solid and the dashed lines according to the value of 
orientation angle, $\theta$. 
The solid line in the right panel of Fig. 5 shows 
the fusion cross sections obtained by averaging the contribution of all 
the orientation angles through Eq. (\ref{orientation}). 
Since the tunneling probability has an exponentially strong dependence 
on the barrier height, the fusion cross sections are significantly 
enhanced for those orientations which yield a lower barrier than 
the spherical case. 
It is remarkable that this 
simple calculation accounts well for 
the experimental 
enhancement of fusion cross sections at subbarrier energies. 
Evidently, nuclear structure effects significantly 
enhance fusion cross sections at energies below the Coulomb barrier, 
which make fusion reactions an interesting probe for nuclear structure. 

\subsection{Coupled-channels equations with full angular momentum coupling}

The nuclear structure effects can be taken into account in a more quantal way 
using the coupled-channels method.  
In order to formulate the coupled-channels method, 
consider a collision between two nuclei in the presence of 
the coupling of the relative motion,  
$\vec{r}=(r,\hat{\vec{r}})$, to a
nuclear intrinsic motion $\xi$. 
We assume the following Hamiltonian for this system,  
\begin{equation}
H(\vec{r},\xi)=-\frac{\hbar^2}{2\mu}\nabla^2+V(r)+H_0(\xi)
+V_{\rm coup}(\vec{r},\xi), 
\label{Htot}
\end{equation}
where $H_0(\xi)$ and $V_{\rm coup}(\vec{r},\xi)$ 
are the intrinsic and the coupling Hamiltonians, respectively. 
In general the intrinsic degree of freedom $\xi$ has a finite spin. 
We therefore expand the coupling Hamiltonian in multipoles as 
\begin{equation}
V_{\rm coup}(\vec{r},\xi)
=\sum_{\lambda>0}f_{\lambda}(r)Y_{\lambda}
(\hat{\vec{r}})\cdot
T_{\lambda}(\xi).
\end{equation}
Here $Y_{\lambda}(\hat{\vec{r}})$ are the spherical 
harmonics and $T_{\lambda}(\xi)$ are spherical tensors constructed from
the intrinsic coordinate. The dot indicates a scalar product.
The sum is taken over all values of $\lambda$ except for $\lambda=0$, 
which is already included in the bare potential, $V(r)$.

For a given total angular momentum $J$ and its $z$ component $M$, 
one can define the channel wave functions as 
\begin{equation}
\langle \hat{\vec{r}}\xi\vert(\alpha lI)JM\rangle
=\sum_{m_l,m_I}\langle lm_lIm_I\vert JM\rangle 
Y_{lm_l}(\hat{\vec{r}})\varphi_{\alpha Im_I}(\xi),
\end{equation}
where $l$ and $I$ are the orbital and the intrinsic 
angular momenta, respectively. 
$\varphi_{\alpha Im_I}(\xi)$ are the wave functions of the intrinsic motion 
which obey 
\begin{equation}
H_0(\xi)\varphi_{\alpha Im_I}(\xi)=\epsilon_{\alpha I}\,\varphi_{\alpha Im_I}(\xi).
\end{equation}
Here, $\alpha$ denotes any quantum number besides the angular momentum. 
Expanding the total wave function with the channel wave functions as
\begin{equation}
\Psi_{J}(\vec{r},\xi)=
\sum_{\alpha,l,I}\frac{u^{J}_{\alpha lI}(r)}{r}
\langle \hat{\vec{r}}\xi\vert(\alpha lI)JM\rangle,
\end{equation}
the coupled-channels equations for $u^{J}_{\alpha lI}(r)$ read
\begin{equation}
\left[-\frac{\hbar^2}{2\mu}\frac{d^2}{dr^2}
+\frac{l(l+1)\hbar^2}{2\mu r^2}+V(r)
-E+\epsilon_{\alpha I}\right]u^{J}_{\alpha lI}(r)
+\sum_{\alpha',l',I'}V^{J}_{\alpha lI; \alpha'l'I'}(r)u^{J}_{\alpha'l'I'}(r)=0,
\label{cc0}
\end{equation}
where the coupling matrix elements $V^{J}_{\alpha lI ; \alpha'l'I'}(r)$ 
are given as \cite{Edmonds}
\begin{eqnarray}
V^{J}_{\alpha lI ; \alpha'l'I'}(r)
&=&\langle (\alpha lI)JM\vert V_{\rm coup}(\vec{r},\xi)\vert 
(\alpha'l'I')JM\rangle, \\
&=&\sum_{\lambda}(-)^{I-I'+l'+J}
f_{\lambda}(r)\langle l||Y_{\lambda}||l'\rangle\langle 
\alpha I||T_{\lambda}||\alpha'I'\rangle \nonumber \\
&& \times \sqrt{(2l+1)(2I+1)}\left\{
\begin{array}{ccc}
I'&l'&J \\
l&I &\lambda
\end{array}\right\}. 
\end{eqnarray}
Notice that these matrix elements are independent of $M$. 

For the sake of simplicity of the notation, in the following let us 
introduce a simplified notation, $n=\{\alpha,l,I\}$, and suppress 
the index $J$. The coupled-channels equation (\ref{cc0}) then reads,
\begin{equation}
\left[-\frac{\hbar^2}{2\mu}\frac{d^2}{dr^2}
+\frac{l_n(l_n+1)\hbar^2}{2\mu r^2}+V(r)
-E+\epsilon_n\right]u_n(r)
+\sum_{n'}V_{nn'}(r)u_{n'}(r)=0. 
\label{cc}
\end{equation}
These coupled-channels equations are solved with the incoming wave 
boundary conditions of 
\begin{eqnarray}
u_n(r)&\sim&
\sqrt{\frac{k_{n_i}}{k_n(r)}}\,{\cal T}^J_{nn_i}
\exp\left(-i\int^r_{r_{\rm abs}}k_n(r')dr'\right) 
~~~~~~~~~~~~~~~~r \leq r_{\rm abs}, \\
&=& H_{l_n}^{(-)}(k_nr)\delta_{n,n_i}
-\sqrt{\frac{k_{n_i}}{k_n}}\,{\cal S}^J_{nn_i}H_{l_n}^{(+)}(k_nr)
~~~~~~~~~~~~~~r\to\infty,
\end{eqnarray}
where $n_i$ denotes the entrance channel. 
The local 
wave number $k_n(r)$ is defined by 
\begin{equation}
k_n(r)=\sqrt{
\frac{2\mu}{\hbar^2}\left(E-\epsilon_n-\frac{l_n(l_n+1)\hbar^2}{2\mu r^2}
-V(r)\right)}, 
\label{localk}
\end{equation}
whereas $k_n=k_n(r=\infty)=\sqrt{2\mu(E-\epsilon_n)/\hbar^2}$. 
Once the transmission coefficients ${\cal T}^J_{nn_i}$ are obtained, 
the inclusive penetrability of the Coulomb potential barrier is 
given by
\begin{equation}
P_J(E)=
\sum_{n}
\vert {\cal T}^J_{nn_i} \vert ^2. 
\label{Pcc}
\end{equation}
The fusion cross section is then given by
\begin{equation}
\sigma_{\rm fus}(E)=\frac{\pi}{k^2}\sum_{J}(2J+1)P_{J}(E),
\label{fusion}
\end{equation}
where we have assumed that the initial intrinsic state 
has spin zero, $I_i=0$. 
This equation for fusion cross section 
is similar to Eq. (\ref{fuscross2}) 
except that the penetrability $P_J(E)$ is now 
influenced by the channel coupling effects. 

\subsection{Iso-centrifugal approximation}

The full coupled-channels calculations (\ref{cc}) quickly become 
intricate if many physical channels are included. 
The dimension of the resulting coupled-channels problem is in general 
too large for practical purposes. 
For this reason, the iso-centrifugal approximation, 
which is sometimes referred to as 
the no-Coriolis approximation or the rotating frame approximation, 
has often been introduced 
\cite{NBT86,HRK99,HR04,TI86,T87,ELP87,KRNR93,GCJ86}. 
In the iso-centrifugal approximation to the coupled-channels equations, 
Eq. (\ref{cc}), 
one first 
replaces the angular momentum of the relative motion in each channel 
by the total angular momentum $J$, that is,  
\begin{equation}
\frac{l_n(l_n+1)\hbar^2}{2\mu r^2}\approx \frac{J(J+1)\hbar^2}{2\mu r^2}.
\end{equation}
This corresponds to assuming that the change in the orbital 
angular momentum due to the excitation 
of the intrinsic degree of freedom is negligible. 
Introducing the weighted average wave function 
\begin{equation}
\bar{u}_{I}(r)=(-)^{I}\sum_{l}\langle J 0 I 0|l 0\rangle u_{lI}(r),
\end{equation}
where we have suppressed the index $\alpha$ for simplicity, 
and using the relation
\begin{equation}
\sum_l(-)^{l'+J+\lambda}\sqrt{2l+1}
\left\{
\begin{array}{ccc}
J&I&l \\
\lambda & l' & I'\\
\end{array}
\right\}
\langle l 0 \lambda 0|l' 0\rangle \langle J 0 I 0|l 0\rangle 
=\frac{(-)^{I'}}{\sqrt{2I+1}}
\langle J 0 I' 0|l' 0\rangle \langle I' 0 \lambda 0|I 0\rangle, 
\end{equation}
one finds that 
the wave function $\bar{u}_{I}(r)$ obeys  
the reduced coupled-channels equations,
\begin{equation}
\left(-\frac{\hbar^2}{2\mu}\frac{d^2}{dr^2}+ 
\frac{J(J+1)\hbar^2}{2\mu r^2}+V(r)-E+\epsilon_{I}\right)
\bar{u}_{I}(r)
+\sum_{I'}\sum_\lambda
\sqrt{\frac{2\lambda+1}{4\pi}}f_\lambda(r)
\langle\varphi_{I0}|T_{\lambda 0}|\varphi_{I'0}\rangle 
\bar{u}_{I'}(r)=0. 
\label{ccisocentrifugal}
\end{equation}
These are nothing but the coupled-channels equations for a spin-zero 
system with the interaction Hamiltonian given by 
\begin{equation}
V_{\rm coup}=
\sum_\lambda f_\lambda(r)Y_\lambda(\hat{\vec{r}}=0)\cdot T_{\lambda}
=\sum_\lambda\sqrt{\frac{2\lambda+1}{4\pi}}f_\lambda(r)T_{\lambda 0}.
\label{rotatingframe}
\end{equation}
In solving the reduced coupled-channels equations, 
similar boundary conditions are imposed for $\bar{u}_{I}$ as those for 
$u_{lI}$, 
\begin{eqnarray}
\bar{u}_{I}(r)&\sim&
\sqrt{\frac{k_{I_i}}{k_{I}(r)}}\,\bar{{\cal T}}^J_{II_i}
\exp\left(-i\int^r_{r_{\rm abs}}k_I(r')dr'\right) 
~~~~~~~~~~~~~~~~r \leq r_{\rm abs}, \\
&=& H_{J}^{(-)}(k_Ir)\delta_{I,I_i}
-\sqrt{\frac{k_{I_i}}{k_I}}\,{\cal \bar{S}}^J_{II_i}H_{J}^{(+)}(k_Ir)
~~~~~~~~~~~~~~r\to\infty,
\end{eqnarray}
where $k_I$ and $k_I(r)$ are defined 
in the same way 
as in Eq. (\ref{localk}). 
The fusion cross section is then given by Eq. (\ref{fusion}) with the 
penetrability of
\begin{equation}
P_{J}(E)=
\sum_{I}
\vert \bar{{\cal T}}^J_{II_i} \vert ^2.
\end{equation}

Since the reduced coupled-channels equations in the iso-centrifugal 
approximation are equivalent to the coupled-channels equations with 
a spin-zero intrinsic motion, 
the complicated angular momentum couplings disappear. 
A remarkable fact is that the dimension of the coupled-channels 
equations is drastically reduced in this approximation. 
For example, 
if one includes four intrinsic states with 2$^+$, 4$^+$, 
6$^+$, and 8$^+$ together with the ground state 
in the coupled-channels equations, the original 
coupled-channels  
have 25 dimensions for $J \geq 8$, 
while 
the dimension is reduced to 5 in the iso-centrifugal approximation. 
The validity of the iso-centrifugal approximation 
has been well tested for heavy-ion fusion 
reactions, 
and it has been concluded that the iso-centrifugal 
approximation leads to negligible errors in 
calculating fusion cross sections \cite{T87,HR04}. 

\subsection{Coupling to low-lying collective states}

\subsubsection{Vibrational coupling}

Let us now discuss the explicit form of the coupling Hamiltonian $V_{\rm coup}$ 
for heavy-ion fusion reactions. We first consider couplings of 
the relative motion to the 2$^{\lambda}$-pole surface vibration of a target 
nucleus. 
In the geometrical model of Bohr and Mottelson, the radius of the 
vibrating target is parameterized as 
\begin{equation}
R(\theta,\phi) = R_T \left(1+
\sum_{\mu}\alpha_{\lambda\mu}
Y_{\lambda\mu}^*(\theta,\phi) 
\right),
\end{equation}
where $R_T$ is the equivalent sharp surface radius and $\alpha_{\lambda\mu}$ 
is the surface coordinate of the target nucleus.  To the lowest order, 
the surface oscillation is approximated by a harmonic oscillator and 
the Hamiltonian for the intrinsic motion is given by 
\begin{equation}
H_0=\hbar\omega_\lambda \left(\sum_{\mu} a_{\lambda\mu}^{\dagger}a_{\lambda\mu}
+\frac{2\lambda+1}{2}\right). 
\end{equation}
Here $\hbar\omega_\lambda$ is the oscillator quanta 
and $a_{\lambda\mu}^{\dagger}$ 
and $a_{\lambda\mu}$ are the phonon creation and annihilation operators, 
respectively. The surface coordinate $\alpha_{\lambda\mu}$ is related to the 
phonon creation and annihilation operators by 
\begin{equation}
\alpha_{\lambda\mu}=
\alpha_0 \left(a_{\lambda\mu}^{\dagger}+(-)^{\mu}
a_{\lambda\mu} \right)
=\frac{\beta_\lambda}{\sqrt{2\lambda+1}}\, 
\left(a_{\lambda\mu}^{\dagger}+(-)^{\mu}
a_{\lambda\mu} \right), 
\end{equation}
where 
$\alpha_0=\beta_{\lambda}/\sqrt{2\lambda+1}$ 
is the amplitude of the zero point motion\cite{BM75}. 
The deformation parameter $\beta_\lambda$ 
can be estimated 
from the experimental transition probability using 
(see Eq. (\ref{multipole}) below) 
\begin{equation}
\beta_\lambda=\frac{4\pi}{3Z_TR_T^{\lambda}}
\sqrt{\frac{B(E\lambda)\uparrow}{e^{2}}}. 
\end{equation}

The surface vibration of the target nucleus 
modifies both the nuclear and the Coulomb interactions 
between the colliding nuclei. 
In the collective model, the nuclear interaction is assumed to be 
a function of the separation distance between the vibrating surfaces 
of the colliding nuclei, and thus is given as 
\begin{equation}
V^{(N)}(\vec{r},\alpha_{\lambda\mu})
=V_N\left(r-R_T
\sum_{\mu}\alpha_{\lambda\mu}
Y_{\lambda\mu}^*(\hat{\vec{r}}) \right).
\label{nucl-full}
\end{equation}
If the amplitude of the zero point motion of the vibration is small, 
one can expand 
this equation in terms of $\alpha_{\lambda\mu}$ and keep only 
the linear term, 
\begin{equation}
V^{(N)}(\vec{r},\alpha_{\lambda\mu})
=V_N(r)-R_T\frac{dV_N(r)}{dr}\sum_{\mu}\alpha_{\lambda\mu}
Y_{\lambda\mu}^*(\hat{\vec{r}}). 
\label{coupN}
\end{equation}
This approximation is called the linear coupling approximation. 
The first term of the right hand side (r.h.s.) of Eq. (\ref{coupN}) 
is the bare nuclear potential in the absence of 
the coupling, while the second 
term is the nuclear component of the coupling Hamiltonian. 
Even though the linear coupling 
approximation does not work well for heavy-ion fusion 
reactions\cite{HRK99,HTDHL97}, 
we employ it in this subsection in order to illustrate the 
coupling scheme. 
In Sec. 3.5, we will discuss how the higher order terms can 
be taken into account in the coupling matrix. 

The Coulomb component of the coupling Hamiltonian is evaluated 
as follows. 
The Coulomb potential 
between the spherical projectile and the vibrating target is given by 
\begin{equation}
V_C(\vec{r})=\int d\vec{r}' 
\frac{Z_PZ_Te^2}{|\vec{r}-\vec{r}'|}
\rho_T(\vec{r}') 
=\frac{Z_PZ_Te^2}{r}+\sum_{\lambda'\ne 0}\sum_{\mu'}
\frac{4\pi Z_Pe}{2\lambda'+1}
Q_{\lambda'\mu'}Y_{\lambda'\mu'}^*(\hat{\vec{r}})\frac{1}{r^{\lambda'+1}}, 
\label{coupC}
\end{equation}
where $\rho_T$ is the charge density of the target nucleus 
and $Q_{\lambda'\mu'}$ the electric multipole operator defined by 
\begin{equation}
Q_{\lambda'\mu'}=\int d\vec{r} Z_Te\rho_T(\vec{r})
r^{\lambda'}Y_{\lambda'\mu'}(\hat{\vec{r}}). 
\end{equation}
The first term of the r.h.s. of Eq. (\ref{coupC}) 
is the bare Coulomb interaction, 
and the second term is the Coulomb component of the coupling Hamiltonian. 
In obtaining Eq. (\ref{coupC}), we have used the formula 
\begin{equation}
\frac{1}{|\vec{r}-\vec{r}'|}=\sum_{\lambda'\mu'}
\frac{4\pi}{2\lambda'+1}\frac{r^{\lambda'}_<}{r^{\lambda'+1}_>}
Y_{\lambda'\mu'}(\hat{\vec{r}}')
Y_{\lambda'\mu'}^*(\hat{\vec{r}}), 
\end{equation}
and have assumed that the relative coordinate $r$ is larger than the charge 
radius of the target nucleus. 
If we assume a sharp matter distribution for the target nucleus, 
the electric multipole operator is given by 
\begin{equation}
Q_{\lambda'\mu'}=\frac{3e}{4\pi}Z_T R_T^{\lambda}
\alpha_{\lambda\mu}\delta_{\lambda\mu,\lambda'\mu'},
\label{multipole}
\end{equation}
up to the first order in the surface coordinate 
$\alpha_{\lambda\mu}$. 

By combining Eqs. (\ref{coupN}), (\ref{coupC}), and (\ref{multipole}), the 
coupling Hamiltonian is expressed by 
\begin{equation}
V_{\rm coup}(\vec{r},\alpha_{\lambda}) = f_{\lambda}(r) 
\sum_{\mu}\alpha_{\lambda\mu}
Y_{\lambda\mu}^*(\hat{\vec{r}}), 
\label{vcoup}
\end{equation}
up to the first order of $\alpha_{\lambda\mu}$. 
Here, $f_{\lambda}(r)$ is the coupling form factor, given by 
\begin{equation}
f_{\lambda}(r)=-R_T\frac{dV_N}{dr}
+\frac{3}{2\lambda+3}Z_PZ_Te^2\frac{R_T^{\lambda}}
{r^{\lambda+1}}, 
\end{equation}
where the first and the second terms are the nuclear and the Coulomb 
coupling form factors, respectively. 
Transforming to the rotating frame, 
the coupling Hamiltonian used in the iso-centrifugal approximation 
is then given by (see Eq. (\ref{rotatingframe})), 
\begin{equation}
V_{\rm coup}(r,\alpha_{\lambda 0}) = \sqrt{\frac{2\lambda+1}{4\pi}} 
f_{\lambda}(r) \alpha_{\lambda0} = 
\frac{\beta_{\lambda}}{\sqrt{4\pi}}f_{\lambda}(r)
\left(a_{\lambda 0}^{\dagger}+a_{\lambda 0}\right). 
\label{coupvib}
\end{equation}
Notice that the coupling form factor $f_{\lambda}$ has the value  
\begin{equation}
f_{\lambda}(R_b)=\frac{Z_PZ_Te^2}{R_b}
\left(\frac{3}{2\lambda+3}\frac{R_T^{\lambda}}
{R_b^{\lambda}}-\frac{R_T}{R_b}\right). 
\end{equation}
at the position of the bare Coulomb barrier, $R_b$, and 
the coupling strength is approximately proportional to the charge product 
of the colliding nuclei. 

In the previous subsection, we showed that the iso-centrifugal 
approximation 
drastically reduces the dimension of the coupled-channels 
equations. 
A further reduction can be achieved by 
introducing effective multi-phonon channels \cite{TI86,KRNR93}.  
In general, the multi-phonon states of the vibrator have several 
levels, which are distinguished from each other 
by the angular momentum and the seniority \cite{BM75}. 
For example, for the 
quadrupole surface vibrations, the two-phonon state has three levels 
($0^+,2^+,4^+$), which are 
degenerate in energy in the harmonic limit.  
The one-phonon state, $|2_1^+\rangle=a_{20}^\dagger|0\rangle$, 
couples only to a particular combination of 
these triplet states, 
\begin{equation}
|2\rangle = \sum_{I=0,2,4} \langle 2020|I0\rangle |I0\rangle = 
\frac{1}{\sqrt{2!}}(a^{\dagger}_{20})^2 |0\rangle. 
\end{equation}
It is thus sufficient to include this single state in the calculations, 
instead of 
three triplet states. 
In the same way, one can introduce the $n$-phonon channel 
for a multipolarity $\lambda$ as 
\begin{equation}
|n\rangle = \frac{1}{\sqrt{n!}}(a^{\dagger}_{\lambda 0})^n |0\rangle. 
\end{equation}
See Appendix C for the case of two different vibrational modes 
of excitation ({\it e.g.,} a quadrupole and an octupole vibrations). 

If one truncates the phonon space up to the two-phonon state, 
the corresponding coupling 
matrix is then given by 
\begin{equation}
V_{\rm coup}=
\left(\begin{array}{ccc}
0&F(r)&0\\
F(r)&\hbar\omega_\lambda
&\sqrt{2}F(r)\\
0&\sqrt{2}F(r)&2\hbar\omega_\lambda
\end{array}\right),
\label{vib}
\end{equation}
where $F(r)$ is defined as $\beta_{\lambda} f_{\lambda}(r)/\sqrt{4\pi}$. 

The effects of 
deviations from the harmonic oscillator limit presented in this 
subsection 
on subbarrier fusion reactions 
have been discussed in Refs. \citen{HTK97} and \citen{HKT98}. 

\subsubsection{Rotational coupling}

We next consider couplings to the ground rotational band of a 
deformed target. 
To this end, it is convenient to transform to the body 
fixed frame where the $z$ axis is along the orientation of 
the deformed target. The surface coordinate $\alpha_{\lambda\mu}$ 
is then transformed to 
\begin{equation}
a_{\lambda\mu}=\sum_{\mu'}D^{\lambda}_{\mu'\mu}(\phi_d,\theta_d,\chi_d)
\alpha_{\lambda\mu'},
\end{equation}
where $\phi_d, \theta_d$, and $\chi_d$ are the Euler angles which specify 
the body-fixed frame, thus 
the orientation of the target. If we are particularly interested 
in the quadrupole deformation ($\lambda$=2), the surface coordinates 
in the body fixed frame are expressed as 
\begin{eqnarray}
a_{20}&=&\beta_2\cos\gamma, \\
a_{22}&=&a_{2-2}=\frac{1}{\sqrt{2}}\beta_2\sin\gamma, \\
a_{21}&=&a_{2-1}=0. 
\end{eqnarray}
If we further assume that the deformation is axial symmetric ({\it i.e.}, 
$\gamma=0$), 
the coupling Hamiltonian for the rotational coupling reads (see Eq. 
(\ref{vcoup})) 
\begin{equation}
V_{\rm coup}(\vec{r},\theta_d,\phi_d) = f_{2}(r) 
\sum_{\mu}\beta_2\sqrt{\frac{4\pi}{5}}Y_{2\mu}(\theta_d,\phi_d)
Y_{2\mu}^*(\hat{\vec{r}}). 
\end{equation}
In order to obtain this equation, we have used the relation 
\begin{equation}
D^{L}_{M0}(\phi,\theta,\chi)=\sqrt{\frac{4\pi}{2L+1}}Y_{LM}^*(\theta,\phi). 
\end{equation}
The coupling Hamiltonian in the rotating frame is thus given by 
\begin{equation}
V_{\rm coup}(r,\theta) = f_{2}(r) 
\beta_2Y_{20}(\theta),
\label{couprot}
\end{equation}
where $\theta$ is the angle between $(\theta_d,\phi_d)$ and 
$\hat{\vec{r}}$, that is, the direction of the orientation 
of the target measured from the direction of the relative motion 
between the colliding nuclei. 
Since the wave function for the $|I0\rangle$ state in the ground rotational 
band is given by $|I0\rangle = |Y_{I0}\rangle$, 
the corresponding coupling matrix is given by 
\begin{equation}
V_{\rm coup}=f_2(r)\beta_2\langle Y_{I'0}|Y_{20}|Y_{I0}\rangle 
=\left(\begin{array}{ccc}
0&F(r)&0\\
F(r)&\epsilon_2+\frac{2\sqrt{5}}{7}F(r)&\frac{6}{7}F(r)\\
0&\frac{6}{7}F(r)&\frac{10}{3}\epsilon_2+\frac{20\sqrt{5}}{77}F(r)
\end{array}\right),
\label{rot}
\end{equation}
when the rotational band is truncated at the first 4$^+$ state. 
Here, $\epsilon_2$ is the excitation energy of the 
first 2$^+$ state, and $F(r)$ is defined as $\beta_{2} f_{2}(r)/\sqrt{4\pi}$ 
as in Eq. (\ref{vib}). 

One of the main differences between the vibrational (\ref{vib}) and 
the rotational (\ref{rot}) couplings is that the latter has a diagonal 
component which is proportional to the deformation parameter $\beta_2$. 
The diagonal component in the rotational coupling 
is referred to as the {\it reorientation effect} and 
has been used in the Coulomb excitation technique to determine the 
sign of the deformation parameter \cite{BE68}. 
Notice that the results of the coupled-channels calculations are independent 
of the sign of $\beta_2$ for the vibrational coupling. 

The effects of the $\gamma$ deformation on subbarrier fusion were studied 
in Ref. \citen{DFV97}. If there is a finite $\gamma$ deformation, the 
coupling Hamiltonian in the rotating frame becomes 
\begin{equation}
V_{\rm coup}(r,\theta,\phi) = f_{2}(r) 
\left(\beta_2\cos\gamma Y_{20}(\theta) +\frac{1}{\sqrt{2}}\beta_2\sin\gamma 
\left(Y_{22}(\theta,\phi)+Y_{2-2}(\theta,\phi)\right)\right).
\end{equation}
Higher order deformations can also be taken into account in a similar way 
as the quadrupole deformation. 
For example, if there is an axial symmetric hexadecapole 
deformation in addition to quadrupole deformation, the 
coupling Hamiltonian reads 
\begin{equation}
V_{\rm coup}(r,\theta) = f_{2}(r) \beta_2 Y_{20}(\theta) + 
f_{4}(r) \beta_4 Y_{40}(\theta),
\end{equation}
where $\beta_4$ is the hexadecapole deformation parameter. 

\subsection{All order couplings}

In the previous subsection, for simplicity, 
we have used the linear coupling approximation 
and expanded the coupling Hamiltonian in terms of the deformation parameter. 
However, it has been shown that the higher order terms play an important 
role in heavy-ion subbarrier fusion reactions 
\cite{HRK99,HTDHL97,EL87,BBK93,R95,SACN95}. 
These higher order terms can be evaluated as follows\cite{HRK99}. 
If we employ the Woods-Saxon potential, Eq. (\ref{WS}), 
the nuclear coupling Hamiltonian can be generated by changing the
target radius in the potential to a dynamical
operator
\begin{equation}
R_0 \to R_0 + \hat{O}, 
\end{equation}
that is, 
\begin{equation}
V_N(r)\to V_N(r,\hat{O})= -\frac{V_0}{1+\exp((r-R_0-\hat{O})/a)}.
\label{Ncoup}
\end{equation}
For the vibrational coupling, the operator 
$\hat{O}$ is given by (see Eq. (\ref{coupvib})), 
\begin{equation}
\hat{O}=\frac{\beta_{\lambda}}{\sqrt{4\pi}}R_T
(a_{\lambda 0}^{\dagger} + a_{\lambda 0}),
\end{equation}
while for the rotational coupling it is given by 
(see Eqs. (\ref{defWS}) and and (\ref{couprot})), 
\begin{equation}
\hat{O}=\beta_2R_TY_{20}(\theta)+\beta_4R_TY_{40}(\theta).
\end{equation}
The matrix elements of the coupling Hamiltonian 
can be easily obtained using a matrix algebra \cite{KR93}.
In this algebra, one first looks for the eigenvalues and eigenvectors
of the operator $\hat{O}$ which satisfies
\begin{equation}
\hat{O}|\alpha\rangle = \lambda_{\alpha} |\alpha\rangle.
\end{equation}
This is done by numerically diagonalising the matrix
$\hat{O}$, whose elements are given by
\begin{equation}
\hat{O}_{nm}=\frac{\beta_{\lambda}}{\sqrt{4\pi}}R_T
(\sqrt{m}\delta_{n,m-1}+\sqrt{n}\delta_{n,m+1}).
\end{equation}
for the vibrational case, and 
\begin{eqnarray}
\hat{O}_{II'}
&=&
\sqrt{\frac{5(2I+1)(2I'+1)}{4\pi}}\beta_2 R_T
\left(\begin{array}{ccc}
I&2&I'\\
0&0&0
\end{array}\right)^2 \nonumber \\
&&+
\sqrt{\frac{9(2I+1)(2I'+1)}{4\pi}}\beta_4 R_T
\left(\begin{array}{ccc}
I&4&I'\\
0&0&0
\end{array}\right)^2.
\end{eqnarray}
for the rotational case. 
The nuclear coupling matrix elements are then evaluated as
\begin{eqnarray}
V_{nm}^{(N)}&=&\langle n|V_N(r,\hat{O})|m\rangle
-V_N(r)\delta_{n,m}, \nonumber \\
&=&\sum_{\alpha}\langle n|\alpha\rangle\langle\alpha|m\rangle
V_N(r,\lambda_{\alpha})
-V_N(r)\delta_{n,m}.
\end{eqnarray}
The last term in this equation is included to avoid the double counting
of the diagonal component.

The computer code {\tt CCFULL} has been written with this scheme\cite{HRK99}, 
and has been used in analysing recent experimental fusion cross 
sections for many systems. {\tt CCFULL} also includes the second order 
terms in the Coulomb coupling for the rotational case, while it uses 
the linear coupling approximation for the Coulomb coupling in the 
vibrational case\cite{HRK99}. 

\subsection{WKB approximation for multi-channel penetrability}

\begin{figure}[t]
\centerline{\includegraphics[width=11cm,clip]{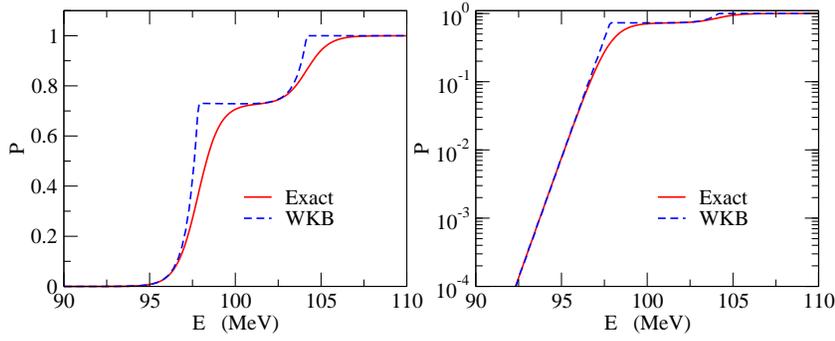}}
\caption{
The barrier penetrability for a two-level problem as a function of 
energy $E$ in the linear (the left panel) and the logarithmic (the right 
panel) scales. 
The solid and the dashed lines are the exact solution and the WKB 
approximation, respectively. }
\end{figure}

Whereas the coupled-channels equations, Eq. (\ref{ccisocentrifugal}), 
can be numerically solved {\it e.g., } with the computer code {\tt CCFULL} 
once the coupling Hamiltonian has been set up, 
it is always useful to have an approximate solution. 
In the next section, we will discuss the 
limit of zero excitation energy 
for intrinsic degrees of freedom, 
in which the coupled-channels equations are decoupled. 
In this subsection, on the other hand, 
we discuss another approximate solution based on the semiclassical 
approximation. 

The penetrability in the WKB approximation is well known for a 
one dimensional potential $V(x)$ and is given by,
\begin{equation}
P(E)=
\exp\left[-2\int^{x_1}_{x_0}dx'\sqrt{\frac{2\mu}{\hbar^2}
(V(x')-E)}\right], 
\label{wkb0}
\end{equation}
where $x_0$ and $x_1$ are the inner and the outer turning points 
satisfying $V(x_0)=V(x_1)=E$, respectively. 
One can also introduce the uniform approximation 
to take into account the multiple 
reflection under the barrier, 
and obtain a formula which is valid 
at all energies from below to above the barrier
\cite{Brink85,BS83,bt77,ltm78,lt78}, 
\begin{equation}
P(E)=
\frac{1}{1+\exp\left[2\int^{x_1}_{x_0}dx'\sqrt{\frac{2\mu}{\hbar^2}
(V(x')-E)}\right]}.  
\end{equation}
It has been shown in Ref. \citen{HB04} that 
one can generalize the primitive WKB formula (\ref{wkb0}) 
to a multi-channel 
problem as, 
\begin{equation}
P=\sum_n\left|\left\langle n\left|
\prod_i e^{i\vec{q}(x_i)\Delta x}\right| n_i\right\rangle\right|^2, 
\label{PWKB}
\end{equation}
where 
$\vec{q}(x)=[2\mu(E-\vec{W}(x))/\hbar^2]^{1/2}$ with 
$W_{nm}(x)=\langle n|V(x)+H_0(\xi)+V_{\rm coup}(x,\xi)|m\rangle$ 
(see Eq. (\ref{Htot})). 
Here we have discretized the coordinate $x$ with a mesh spacing 
of $\Delta x$. 
For a single channel problem, Eq. (\ref{PWKB}) is reduced to 
Eq. (\ref{wkb0}). 

Figure 6 shows the result of the multi-channel WKB approximation 
for a two-level problem given by 
\begin{equation}
\vec{W}(x)=\left(
\begin{array}{cc}
V(x)&F(x)\\
F(x)&V(x)+\epsilon
\end{array}
\right) 
=V(x)
\left(
\begin{array}{cc}
1&0\\
0&1
\end{array}
\right)+F(x)
\left(
\begin{array}{cc}
0&1\\
1&0
\end{array}
\right)+
\left(
\begin{array}{cc}
0&0\\
0&\epsilon
\end{array}
\right),
\label{cc2ch}
\end{equation}
with 
\begin{equation}
V(x)=V_0e^{-x^2/2s^2},~~~~~F(x)=F_0e^{-x^2/2s_f^2}. 
\end{equation}
The parameters are chosen following Ref. \citen{DLW83} to be 
$V_0$=100 MeV, $F_0$=3 MeV, and $s=s_f=$3 fm, which mimic 
the fusion reaction between two $^{58}$Ni nuclei. 
The excitation energy $\epsilon$ and the mass $\mu$ are taken to be 
2 MeV and 29$m_N$, respectively, 
where $m_N$ is the nucleon mass. 
It is remarkable that the WKB formula (\ref{PWKB})
reproduces almost perfectly the exact solution 
at energies well below the barrier. 
The WKB formula breaks down at energies around the barrier, as in 
the single-channel problem. 

The figure also suggests that 
the penetrability is given by a weighted sum of 
two penetrabilities, 
\begin{equation}
P(E)=w_1P(E;\lambda_1(x)) +w_2P(E;\lambda_2(x)),
\end{equation}
where $\lambda_i(x)$ are the eigen-potentials, 
$\lambda_i(x)=V(x)+[\epsilon\pm\sqrt{\epsilon^2+4F(x)^2}]/2$, 
obtained by diagonalizing the matrix $\vec{W}(x)$, (\ref{cc2ch}). 
We will discuss this point in the next section. 

\section{Barrier distribution representation of 
multi-channel penetrability}

\subsection{Sudden tunneling limit and barrier distribution}

In the limit of vanishing excitation energy 
for the intrinsic motion ({\it i.e.,}
in the limit of $\epsilon_I\to 0$), 
the reduced coupled-channels
equations (\ref{ccisocentrifugal}) are completely decoupled. 
This limit corresponds to the case where the tunneling occurs much faster 
than the intrinsic motion, and thus is referred to as the sudden 
tunneling limit. 
In this limit,
the coupling matrix defined as 
\begin{equation}
V_{II'}\equiv \epsilon_I\delta_{I,I'} 
+\sqrt{\frac{2\lambda+1}{4\pi}}f_\lambda(r)
\langle\varphi_{I0}|T_{\lambda 0}|\varphi_{I'0}\rangle 
\label{coup}
\end{equation}
can be diagonalized independently of $r$ (for simplicity we consider 
only a single value of $\lambda$). 
See also Eq. (\ref{cc2ch}). 
It is then easy to prove that the fusion cross section 
is given as a weighted sum of the cross sections 
for uncoupled eigenchannels\cite{NBT86,NRL86}, 
\begin{equation}
\sigma_{\rm fus}(E)=\sum_\alpha w_\alpha 
\,\sigma_{\rm fus}^{(\alpha)}(E), 
\label{crossfus}
\end{equation}
where $\sigma_{\rm fus}^{(\alpha)}(E)$ is 
the fusion cross section for a potential 
in the eigenchannel $\alpha$, {\it i.e.,}
$V_\alpha(r)=V(r)+\lambda_\alpha(r)$. 
The same relation holds also for the quasi-elastic scattering
\cite{HR04,NRL86,ARN88}. 
Here, 
$\lambda_\alpha(r)$ is the eigenvalue of the coupling matrix
(\ref{coup}) (when $\epsilon_I$ is zero, $\lambda_\alpha(r)$ is 
simply given by $\lambda_\alpha\cdot f_\lambda(r)$). 
The weight factor $w_\alpha$ is given by 
$w_\alpha=|U_{0\alpha}|^2$, where $U$ is the unitary matrix which diagonalizes
Eq. (\ref{coup}). 
Note that the unitarity of the matrix $U$ leads to the 
relation that the sum of all the weight factors, $\sum_\alpha w_\alpha$, is 
unity \cite{NBT86}. 

The resultant formula (\ref{crossfus}) in the sudden tunneling limit 
can be interpreted in the following way. 
In the absence of the coupling, the incident particle 
encounters only the single potential barrier, $V(r)$. 
When the coupling is turned on, the bare potential splits 
into many barriers. Some of them are lower than the 
bare potential and some of them higher. 
In this picture, the potential barriers are distributed 
with appropriate weight factors, $w_\alpha$. 

The orientation average formula discussed in Sec. 3.1 
(see Eq. (\ref{orientation})) for a deformed 
target nucleus can also be obtained from the coupled-channels equations 
by taking the sudden tunneling limit\cite{NBT86}. 
To show this, first notice that the coupling Hamiltonian 
is diagonal with respect to the orientation angle, $\theta$. 
If all the members of the rotational band are included 
in the coupled-channels equations, the eigenstates of the 
coupling Hamiltonian matrix then become the same as the 
angle vector $|\theta\rangle$ with the eigenvalue given by the 
deformed Woods-Saxon potential, Eq.(\ref{defWS})\cite{NBT86,TAB92,HTBB95}. 
The weight factor in this case is simply given by $w(\theta)=
|\langle \theta|\varphi_{I=0}\rangle|^2=|Y_{00}(\theta)|^2$. 

The physical interpretation of the orientation average formula is that 
fusion reaction takes place so suddenly 
that the orientation 
angle is fixed during the fusion reaction. 
This is justified because the first 2$^+$ state of a heavy deformed nucleus 
is small (see Fig. 4), 
corresponding to a large moment of inertia for the rotational 
motion. As the orientation angles are distributed according to the 
wave function for the ground state, 
the fusion cross section can be computed first by fixing the orientation 
angle and then averaging over the orientation angle with the appropriate 
weight factor, $w(\theta)$. 
The applicability of this formula has been investigated in Ref. \citen{RHT01} 
in the reactions of $^{154}$Sm target with various projectiles ranging 
from $^{12}$C to $^{40}$Ar. It has been shown that the formula works well, 
although the agreement with the exact coupled-channels calculations 
which take into account the finite excitation energy 
of the rotational excitation 
becomes slightly worse for a large value of 
charge product of the projectile and the target nuclei. 

\subsection{Fusion barrier distribution}

\begin{figure}[tb]
\centerline{\includegraphics[width=11cm,clip]{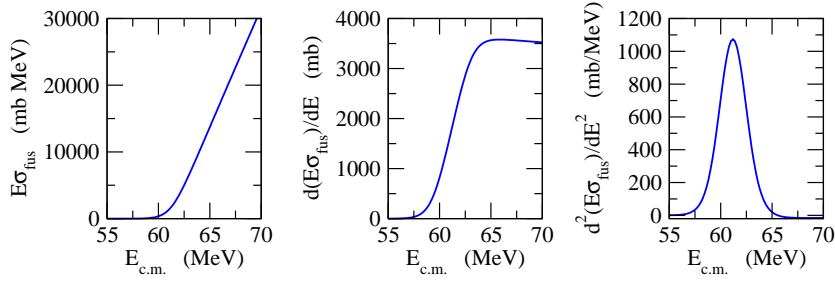}}
\caption{
The product of energy $E$ and the fusion cross section 
$\sigma_{\rm fus}$, $E\sigma_{\rm fus}$, for the $^{16}$O+$^{144}$Sm 
reaction obtained with the potential model (the left panel). 
The middle and the right panels show the first and the second 
energy derivatives 
of $E\sigma_{\rm fus}$, respectively. }
\end{figure}

Rowley, Satchler, and Stelson have proposed 
a method to extract, directly from the experimental fusion cross sections, 
a way how the barriers are distributed \cite{DHRS98,RSS91}. 
In order to illustrate the method, 
let us first discuss the classical fusion cross section given by, 
\begin{equation}
\sigma^{cl}_{\rm fus}(E)=\pi
R_b^2\left(1-\frac{V_b}{E}\right)\,\theta(E-V_b). 
\label{classical}
\end{equation}
From this expression, it is clear that the first derivative of 
$E\sigma^{cl}_{\rm fus}$ is proportional to the classical 
penetrability for a 1-dimensional barrier of height $V_b$, 
\begin{equation}
\frac{d}{dE}[E\sigma^{cl}_{\rm fus}(E)]=\pi R_b^2\,\theta(E-V_b)
=\pi R_b^2\,P_{cl}(E),
\end{equation}
and the second derivative to a delta function, 
\begin{equation}
\frac{d^2}{dE^2}[E\sigma^{cl}_{\rm fus}(E)]=\pi R_b^2\,\delta(E-V_b). 
\label{clfus}
\end{equation}

In quantum mechanics, 
the tunneling effect smears the delta function in Eq. (\ref{clfus}). 
As we have noted in Sec. 2.2, 
an analytic formula for  
the fusion cross section can be obtained if one approximates 
the Coulomb barrier by an inverse parabola, see 
Eq. (\ref{wong}) in Appendix B. 
Again, the 
first derivative of 
$E\sigma_{\rm fus}$ is proportional to the s-wave penetrability for a
parabolic barrier, 
\begin{equation}
\frac{d}{dE}[E\sigma_{\rm fus}(E)]=\pi R_b^2\,
\frac{1}{1+\exp\left[-\frac{2\pi}{\hbar\Omega}(E-V_b)\right]} 
=\pi R_b^2\,P(E), 
\label{Ppara}
\end{equation}
and the second derivative is proportional to the derivative of the $s$-wave 
penetrability, 
\begin{equation}
\frac{d^2}{dE^2}
[E\sigma_{\rm fus}(E)] 
=\pi R_b^2\,\frac{2\pi}{\hbar\Omega}\,\frac{e^x}{(1+e^x)^2} = 
\pi R_b^2 \frac{dP(E)}{dE}. 
\label{bdpara}
\end{equation}
As shown in Fig. 7, this function has the following properties: 
i) it is symmetric around $E=V_b$, ii) it is centered at 
$E=V_b$, iii) 
its integral over $E$ is $\pi R_b^2$, and iv) 
it has a relatively narrow width
of around $\ln(3+\sqrt{8})\hbar\Omega/\pi \sim 0.56 \hbar\Omega$. 

\begin{figure}[t]
\centerline{\includegraphics[width=11cm,clip]{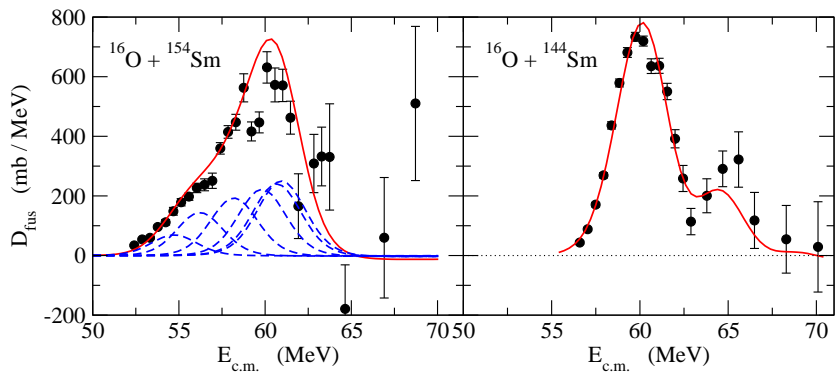}}
\caption{
(The left panel) The fusion barrier distribution 
$D_{\rm fus}(E)=d^2(E\sigma_{\rm fus})/dE^2$ 
for the $^{16}$O+$^{154}$Sm reaction\cite{LDH95}. 
The solid line is obtained with the orientation average formula, 
which corresponds to the solid line in 
Fig.  5. The dashed lines indicate the contributions from six 
individual eigenbarriers ({\it i.e.,} orientation angles).
(The right panel) 
The fusion barrier distribution 
for the $^{16}$O+$^{144}$Sm reaction\cite{LDH95}. 
The solid line shows the result of the coupled-channels calculations which 
take into account the anharmonic double phonon excitations of 
$^{144}$Sm~\cite{HTK97,HKT98}. 
}
\end{figure}

In the presence of channel couplings, Eq. (\ref{crossfus}) immediately 
leads to 
\begin{equation}
D_{\rm fus}=\frac{d^2}{dE^2}[E\sigma_{\rm fus}(E)]=\sum_\alpha w_\alpha 
\,\frac{d^2}{dE^2}[E\sigma_{\rm fus}^{(\alpha)}(E)]. 
\end{equation}
This function has been referred to as fusion barrier distribution. 
As an example,
the left panel of 
Fig. 8 shows 
the barrier distribution 
for the $^{16}$O+$^{154}$Sm reaction, whose fusion cross sections have been 
already shown in Fig. 5. 
We replace the integral in Eq. (\ref{orientation}) with the 
$(I_{\rm max}+2)$-point Gauss quadrature with 
$I_{\rm max}$=10. This corresponds to taking 6 different 
orientation angles\cite{NBT86}. The contributions from each eigenbarrier 
are shown by the dashed line in Fig. 8. 
The solid line is the sum of all the contributions, which is compared 
with the experimental data \cite{LDH95}. 
One can see that the calculation well reproduces the experimental data. 
Moreover, this analysis suggests that $^{154}$Sm is a prolately  
deformed nucleus, since 
if it were an oblate nucleus, 
then lower potential barriers 
would have larger weights, and $D_{\rm fus}$ would be larger 
for smaller $E$, in contradiction to the experimentally  
observed barrier distribution\cite{DHRS98}. 

The fusion barrier distribution
has been extracted for many systems, see Ref. \citen{DHRS98} and references 
therein. 
The extracted barrier distributions 
were shown to be sensitive to the effects of channel-couplings and 
have provided a much more apparent way of understanding their effects on the 
fusion process than the fusion excitation functions themselves. 
These experimental data have thus enabled a detailed study of the effects of 
nuclear intrinsic excitations on fusion reactions, and have generated 
a renewed interest in heavy-ion subbarrier fusion reactions.
An important point is that the nature 
of sub-barrier fusion reactions as a tunneling process 
exponentially amplifies the effects of detailed nuclear structure.
Fusion barrier distribution makes this effect even more visible 
by plotting in the linear scale. 
The sub-barrier fusion reactions thus offer a
novel way of nuclear spectroscopy which could be called 
a tunneling assisted nuclear spectroscopy.
As an example, we mention that it was recently 
applied to elucidate the shape transition and 
the shape coexistence of Ge isotopes\cite{Esbensen03,MMTH10}.
It is worthwhile to mention also 
that the method of the barrier distribution 
has been successfully applied to 
heavy-ion quasi-elastic scattering \cite{HR04,TDH94}. 

\subsection{Eigenchannel representation}

As we have discussed in the previous subsection, 
the barrier distribution 
representation, that is, the second derivative of $E\sigma_{\rm fus}$, 
has a clear 
physical meaning only if the excitation energy of the intrinsic motion 
is zero. The concept holds only approximately when the excitation 
energy is finite. 
Nonetheless, this analysis has been successfully applied to 
systems with relatively large excitation energies\cite{LDH95,MDH94,SACN95}. 
For example, 
the second derivative of $E\sigma_{\rm fus}$ for $^{16}$O + $^{144}$Sm 
fusion reaction 
has a clear double-peaked structure (see the right panel of Fig. 8) 
\cite{LDH95,MDH94}. 
The coupled-channels calculation also yields a similar double-peaked 
structure of the fusion barrier distribution, and 
this structure has been interpreted 
in terms of 
the anharmonic octupole 
phonon excitations in $^{144}$Sm~\cite{HTK97,HKT98}, 
whose excitation energy is 1.8 MeV for 
the first 3$^-$ state. 
Also the analysis of the fusion reaction between $^{58}$Ni and $^{60}$Ni, 
where the excitation energies of quadrupole phonon states 
are 1.45 and 1.33 MeV, respectively, shows that the barrier distribution 
representation depends strongly on the number of phonon states 
included in coupled-channels calculations\cite{SACN95}. 
These analyses suggest that 
the representation of fusion process in terms of the second derivative of 
$E\sigma_{\rm fus}$ is a 
powerful method to study the details of the effects of 
nuclear structure, irrespective of the excitation energy of 
the intrinsic motion. 

When the excitation energy of the intrinsic motion is finite, the 
barrier distribution is defined in terms of the eigen-channels. 
To illustrate it, first notice that Eq. (\ref{Pcc}) can be expressed as 
\begin{equation}
P(E)=(T^\dagger T)_{n_in_i}, 
\end{equation}
using the completeness of the channels $n$ 
(we have suppressed the index $J$). 
We then introduce the eigenfunctions of the Hermitian operator 
$T^\dagger T$ as, 
\begin{equation}
(T^\dagger T)|\phi_k\rangle = \gamma_k|\phi_k\rangle. 
\label{eigen}
\end{equation}
Using this basis, the penetrability is given by 
\begin{equation}
P(E)=\sum_k|\langle\phi_k|n_i\rangle|^2\cdot\gamma_k.
\label{Peigen}
\end{equation}
When the excitation energies $\epsilon_n$ are all zero, as we have discussed 
in Sec. 4.1, one can 
diagonalize the coupling matrix $V_{nn'}(r)$ with the basis set 
which is independent of the radial coordinate $r$. In this case, 
the matrix $T$ 
is diagonal on this basis, and the weight factor 
$|\langle\phi_k|n_i\rangle|^2$ is independent of $E$. 
Eq. (\ref{Peigen}) is a generalization of this scheme, which is applicable 
also when the excitation energies are non-zero. 

\begin{figure}[tb]
\centerline{\includegraphics[width=11cm,clip]{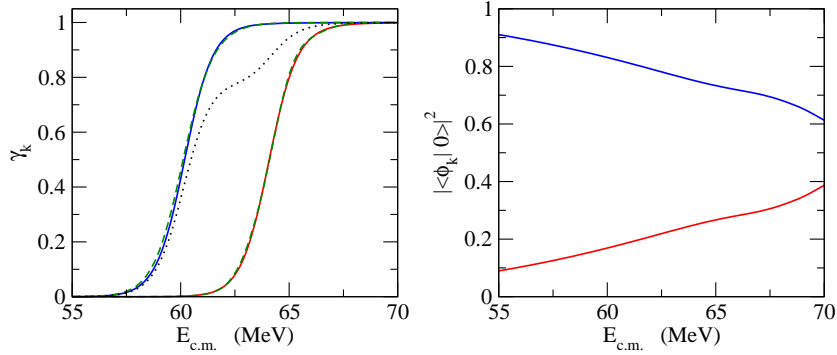}}
\caption{
(The left panel) The $s$-wave penetrabilities for 
the $^{16}$O+$^{144}$Sm reaction. 
The dotted line is obtained with the coupled-channels 
calculations with a single octupole phonon excitation in $^{144}$Sm 
at 1.81 MeV with $\beta_3=0.205$. 
The solid lines show the eigenvalues of the square of the transmission matrix, 
$T^\dagger T$, defined by Eq. (\ref{eigen}). 
The dashed lines denote the penetrabilities of the eigen barriers constructed 
by diagonalizing the coupling matrix at each $r$. 
(The right panel) The weight factors $|\langle\phi_k|n_i\rangle|^2$  
defined in Eq. (\ref{Peigen}) 
as a function of energy. }
\end{figure}

Figure 9 shows the two eigenvalues $\gamma_k$ and the corresponding 
weight factors $|\langle\phi_k|n_i\rangle|^2$ as a function of 
$E$ for a single-phonon coupling calculation for the 
$s$-wave $^{16}$O+$^{144}$Sm 
reaction.  
To this end, we have taken into account couplings to the single 
octupole phonon state in $^{144}$Sm at 1.81 MeV with the deformation 
parameter of $\beta_3$ = 0.205.
The total probability $P(E)$, and the penetrability of the two eigenbarriers, 
obtained by diagonalizing the coupling matrix 
$V_{nn'}(r)$ at each $r$, are also shown in the left panel of the figure 
by the dotted and the dashed lines, respectively. 
One can see that the two eigenvalues $\gamma_k$ approximately 
correspond to the 
penetrability of the eigenbarriers, and thus 
the factors $|\langle\phi_k|n_i\rangle|^2$ can be interpreted as 
the weight factors 
for each eigenbarrier. 
This implies that the fusion cross sections are still given by 
Eq. (\ref{crossfus}) even when the excitation energy is finite, except that 
the eigenbarriers are now constructed by diagonalizing the coupling 
matrix at each $r$. 
The weight factors do not vary strongly 
as a function of energy, suggesting that 
the concept of the fusion barrier
distribution still holds to a good approximation even when the excitation
energy of the intrinsic motion is finite. 
We have reached the same conclusion already in Ref. \citen{HTB97} using 
a different method from the one in this subsection. 
In contrast to the method in Ref. \citen{HTB97}, the method in 
this subsection is more general since the applicability is not restricted to 
a two-level problem. 

\subsection{Adiabatic potential renormalization}

\begin{figure}[tb]
\centerline{\includegraphics[width=12cm,clip]{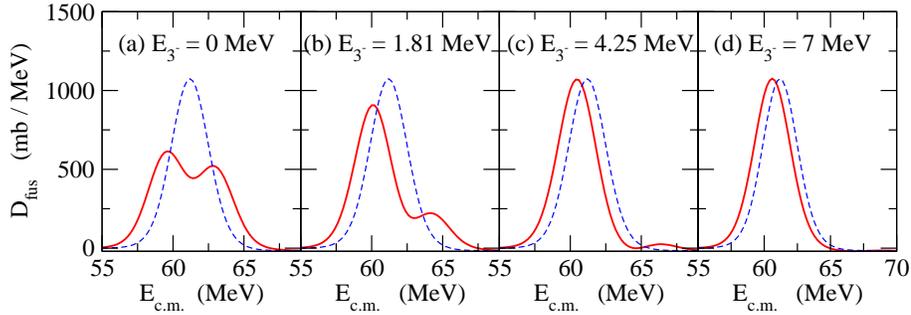}}
\caption{
The fusion barrier distribution $D_{\rm fus}$ for 
the $^{16}$O+$^{144}$Sm reaction with several values of 
excitation energies, $E_{3^-}$, of the octupole vibration in $^{144}$Sm. 
The solid lines are the results of the coupled-channels calculations 
which take into account the single octupole phonon excitation in 
$^{144}$Sm, while the dashed lines are obtained without taking into 
account the channel coupling effect. 
The curvature $\hbar\Omega$ of the Coulomb barrier is 4.25 MeV in these 
calculations.}
\end{figure}

Given that the concept of fusion barrier distribution still holds even 
with a finite excitation energy, it is interesting to investigate 
how the fusion barrier distribution evolves as the excitation energy 
is varied. 
To this end, we carry out the coupled-channels calculations for the 
$^{16}$O+$^{144}$Sm reaction by taking into account the single octupole 
phonon excitation in $^{144}$Sm. 
The solid line in Fig. 10 (a) shows the fusion barrier 
distribution $D_{\rm fus}$ when 
the excitation energy of the octupole vibration, $E_{3^-}$, 
is set to be zero. 
For comparison, the figure also shows the result of no-coupling calculation 
by the dashed line. 
In this case, the original single 
barrier splits into two eigenbarriers with equal weight, 
one corresponds to the effective channel $|0^+\rangle + |3^-\rangle$ and 
the other corresponds to $|0^+\rangle - |3^-\rangle$. 
The fusion barrier distribution is slightly asymmetric since 
the barrier positions, $R_b$, are different between the two effective channels 
(see Eq. (\ref{bdpara})). 

Figure 10(b) corresponds to the physical case of 
$E_{3^-}$ = 1.81 MeV. In this case, the barrier distribution still has 
a clear double peaked structure as in the experimental 
data\cite{LDH95,MDH94}, but the lower energy barrier acquires more weight and 
the barrier distribution is highly asymmetric. 
The effective channels are now $\alpha|0^+\rangle + \beta |3^-\rangle$ 
(the lower energy barrier) 
and $\beta |0^+\rangle - \alpha |3^-\rangle$ (the higher energy barrier) 
with $\alpha > \beta > 0$. 

Figure 10(c) corresponds to the case where the excitation energy is 
set equal to the barrier curvature, $\hbar\Omega$, which is 4.25 MeV 
in the present calculations. 
In this case, the lower energy barrier has an appreciable weight although 
the weight factor for the higher energy barrier is not negligible. 
When the excitation energy is further increased, the weight for the 
lower energy barrier becomes close to unity as is shown in Fig. 10(d), and 
the fusion cross sections are approximately given by 
\begin{equation}
\sigma_{\rm fus}(E)=\sigma_{\rm fus}(E; V(r)+\lambda_0(r)),
\end{equation}
where $V(r)+\lambda_0(r)$ is the lowest eigen-barrier 
(see Eq. (\ref{crossfus})). 
Therefore, the 
main effect of the coupling to a state with a large excitation energy is 
to simply introduce an energy-independent shift of the potential, 
$V(r)\to V(r)+\lambda_0(r)$. This phenomenon is called the adiabatic 
potential renormalization\cite{TMM85,BT85,THAB94}. 
Typical examples in nuclear fusion include the couplings to 
the octupole vibration in $^{16}$O at 6.13 MeV~\cite{HTDHL97b} and to 
giant resonances in general. 

In Refs. \citen{BT85} and \citen{THAB94}, 
it has been argued based on the path integral 
approach to multi-dimensional tunneling that the transition from 
the sudden tunneling to the adiabatic tunneling 
takes place at the excitation energy around the barrier 
curvature, $\hbar\Omega$. That is, if the excitation energy is much larger 
than the barrier curvature, the channel coupling effect can be 
well expressed in terms of the adiabatic barrier renormalization. 
The numerical calculations shown in Fig. 10 are consistent with 
this criterion. 

\section{Fusion at deep subbarrier energies and dissipative tunneling}

Although the coudpled-channels approach has been 
successful for heavy-ion reactions, 
many new challenges have been recognized in recent years. 
One of them is the surface diffuseness anomaly 
discussed in Sec. 2.1. Another challenge, which may also be related to 
the surface diffuseness anomaly\cite{HRD03}, is 
an inhibition of fusion cross sections at deep subbarrier energies. 
This is a phenomenon found only 
recently, when fusion cross
sections could become measured for several systems 
down to extremely low cross sections, 
up to the level of 
few nano barn (nb) \cite{Jiang1,Jiang2,Stefanini08,Montagnoli12,Dasgupta07}.
These experimental data have shown that fusion cross sections
systematically fall off much more steeply at deep-subbarrier energies with
decreasing energy, compared to the expected energy
dependence of cross sections around the Coulomb barrier. That
is, the experimental fusion cross sections appear to be hindered
at deep-subbarrier energies compared to the standard coupled-channels
calculations 
that reproduce the experimental data at subbarrier energies, 
although the fusion cross sections are still
enhanced with respect to a prediction of a single-channel 
potential model. 

\begin{figure}[tb]
\centerline{\includegraphics[width=8cm,clip]{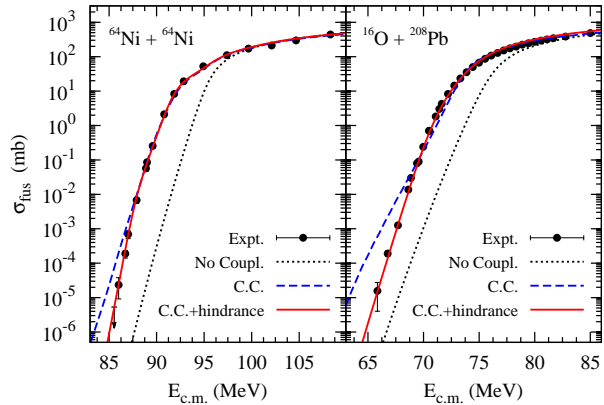}}
\caption{
Fusion cross sections for the $^{64}$Ni+$^{64}$Ni 
and $^{16}$O+$^{208}$Pb systems 
as a function of the incident energy. 
The experimental data are taken from 
Refs. \citen{Jiang1} and \citen{Dasgupta07}. 
The dotted and the dashed lines are the results of 
potential model and the standard coupled-channels calculations, 
respectively. The solid lines denote the result 
when the deep subbarrier fusion hindrance 
 is described in the adiabatic model \cite{IHI09}.}
\end{figure}

Two different models have been proposed so far
in order to account for the deep subbarrier fusion hindrance. 
In the first model, 
assuming the frozen densities in the overlapping
region ({\it i.e.,} the sudden approximation), 
Misicu and Esbensen have introduced a repulsive core 
to an internucleus potential, which is 
originated from the Pauli exclusion principle \cite{ME06}. 
See also Ref. \citen{DP03} for a related publication. 
The resultant potential is 
much shallower than the standard potentials, 
and hinders the fusion
probability for high partial waves. 
In the second model, on the other hand, 
Ichikawa, Hagino, and Iwamoto have 
proposed an adiabatic approach 
by assuming a neck formation 
between the colliding nuclei in the overlap region\cite{IHI07,IHI09}. 
In this model, the reaction is assumed to take place slowly so that the 
density distribution has enough time to adjust to the optimized 
distribution. 
In this adiabatic model, the fusion hindrance originates from the
tunneling of a thick one-body potential due to the neck formation. 
This model has achieved a 
comparable good reproduction of the experimental data to
the sudden model, as is shown in Fig. 11. 

The mechanism for the deep-subbarrier fusion hindrance
has not yet been fully understood, as 
the two different models, in which 
the origin for the deep sub-barrier hindrance is considerably 
different from each other, account for the experimental data equally well. 
However, there is a certain thing which can be concluded by analyzing 
the threshold behavior in deep subbarrier fusion 
\cite{Jiang1,Jiang2,Jiang3,IHI07b,TLH12}, independent of the 
fusion models\cite{IHI07b}. 
In Refs. \citen{Jiang1,Jiang2} and \citen{Jiang3}, the 
deep-subbarrier fusion hindrance has
been analyzed using the astrophysical $S$ factor. It has been
claimed that deep-subbarrier fusion hindrance sets in
at the energy at which the astrophysical $S$ factor reaches its
maximum. The authors of Refs. 
\citen{Jiang1,Jiang2} and \citen{Jiang3} 
even parametrized the threshold energy as a function of charge and mass 
numbers of the projectile and the target nuclei. 
The relation between the
threshold for deep-subbarrier hindrance and the maximum of
the $S$ factor is not clear physically, and thus it is not trivial how
to justify the identification of the threshold energy
with the maximum of the 
astrophysical $S$ factor. Nevertheless, it has turned
out that the threshold energy so obtained closely follows the
values of phenomenological internucleus potentials 
at the touching
configuration \cite{IHI07b}. 
This strongly suggests 
that the dynamics
which takes place after the colliding nuclei touch each other
somehow makes 
the astrophysical $S$ factor decrease as the
incident energy decreases, leading to the fusion hindrance phenomenon. 
Notice that the fusion potential is almost the same between the sudden 
model and the adiabatic model before the touching (see Fig. 1 in 
Ref. \citen{IHI07b}). 

One important aspect of fusion reactions at deep sub-barrier
energies is that the inner turning point of the potential may be
located far inside the touching point of the colliding nuclei (see Fig. 1). 
After the two nuclei touch each other, many non-collective 
excitations of the unified one-body system are activated. 
As is well known from the Caldeira-Leggett model, couplings to 
those excitations lead to energy dissipation, which inhibits 
the tunneling probability\cite{CL81}. 
The energy dissipation may occur also before the touching as a consequence 
of particle transfer processes to 
highly excited states in the target nucleus \cite{Evers11}. 
The phenomenon of deep subbarrier fusion hindrance may therefore be 
a realization of dissipative quantum tunneling, which has been 
extensively studied in many fields of physics and chemistry. 
A characteristic feature in nuclear fusion, which is absent or 
may not be important in dissipative tunneling in other fields, 
is that the couplings to 
(internal) environmental degrees of freedom gradually set in\cite{DH12}. 
That is, before the touching the fully quantum mechanical 
coupled-channels approach 
with couplings to a few collective states of separate nuclei 
is adequate, which however 
gradually loses its validity 
after the touching point 
due to the dissipative couplings\cite{Dasgupta07}. 
This is the region in which the conventional coupled-channels approach does 
not treat explicitly by introducing an absorbing potential or by imposing 
the incoming wave boundary condition. 
Although it is highly important to 
construct a model for 
nuclear fusion by taking into account the dissipative couplings 
\cite{IHI09,DiazTorres08,CCHST86} in order to clarify the deep subbarrier 
fusion hindrance, it is still a challenging open problem. 
To this end, 
a transition from the excitations 
of two separate nuclei in the entrance channel, 
which are included in the 
conventional coupled-channels calculations, 
to {\it molecular} excitations ({\it i.e.,} the excitations of 
the combined mono-nuclear system) 
has to be described in a consistent 
and smooth manner \cite{IHI09,HTS79,MSG81,GMMSS82}. 
The development of quantum mechanical 
version of phenomenological classical 
models for deep inelastic collisions (DIC), such as the wall and window 
formulas for nuclear friction\cite{BBN78,R86,DR87}, 
will also be important in this respect. 

\section{Application of barrier distribution method to surface physics}

The barrier distribution method discussed in Sec. 4 is 
applicable not only to heavy-ion subbarrier fusion reactions but also to 
any multi-channel tunneling problem. 
In general, the barrier distribution is defined as the first derivative 
of penetrability with respect to energy, $dP/dE$
(see Eq. (\ref{bdpara})). 

As an application of the barrier distribution method developed in 
nuclear physics to other fields, let us discuss 
a dissociative adsorption process of diatomic 
molecules on a metal surface. 
When molecular beams are injected on a certain metal, such as Cu and Pd, 
diatomic molecules are broken up 
in the vicinity of 
metal surface 
to two atoms 
due to the molecule-metal interactions 
before they stick to the metal. 
This process is referred to as dissociative 
adsorption, and has been extensively studied in 
surface science together with the inverse process, that is, 
associative desorption \cite{DH95}. 
The 
adsorption process takes place by quantum tunneling at low 
incident energies, 
as 
there is a potential barrier between 
the two phases of the molecules, {\it i.e.,} 
the molecular 
phase and the breakup phase with two separate atoms
\cite{DH95,HSJN94}. 
The vibrational and rotational excitations of diatomic molecules play 
an important role in dissociative adsorption
\cite{RAM92,MRA92,MRAZ93}, as in heavy-ion subbarrier fusion 
reactions. The coupled-channels method has been utilized to discuss the effects 
of the internal excitations of molecules on 
dissociative adsorption\cite{BK89,KO93,CB94,DKO95,DKO96,DKO00,MKD99,GWS95}. 

In this section, we discuss only the simplest case, that is, the 
effect of the rotational excitation on dissociative adsorption, while 
the vibrational degrees of freedom is assumed to be frozen in the 
ground state. In contrast to heavy-ion fusion reactions, the initial rotational 
state 
in the problem of dissociative 
adsorption 
is not necessarily the ground state. 
The initial rotational state of diatomic molecules can be 
in fact selected in molecular beams, and the experimental data of 
Michelsen {\it et al.} \cite{MRA92,MRAZ93} 
have indicated that the adsorption probability of D$_2$ molecules 
on Cu surface 
shows a nonmonotonic behavior as a function of the initial rotational 
state.  
That is, 
at a given incident energy, starting from the initial rotational 
state $L_i=0$, the adsorption probability first decreases for $L_i=5$ 
and then increases for $L_i=10$ and $L_i=14$ (see Fig. 9 in Ref. 
\citen{MRAZ93}). 

In order to explain this behavior, 
Di\~no, Kasai, and Okiji have considered a simple Hamiltonian for 
H$_2$ and D$_2$ molecules given by \cite{DKO95,DKO00}
\begin{equation}
H(s,\theta)=-\frac{\hbar^2}{2M}\frac{\partial^2}{\partial s^2}
+\frac{\hbar^2}{2I(s)}\hat{\vec{L}}^2+V(s,\theta),
\label{Hmolecule}
\end{equation}
where $s$ is the one-dimensional reaction path in the 
two-dimensional potential energy surface spanned by the 
molecule-surface distance, $Z$, and the interatomic distance, $r$. 
The reaction takes place from $s=-\infty$, that corresponds to 
the approaching phase of molecules, to $s=+\infty$, 
where the incident molecule has broken up to two atoms. 
The sticking probability to the metal surface is identified as 
the penetrability of the potential barrier, $V$. 
$M$ in Eq. (\ref{Hmolecule}) 
is the mass for the translational motion of the diatomic molecule 
given by $M=2m$, where $m$ is the mass of the atom ({\it i.e.,}  
$m=m_H$ for H$_2$ molecule and $m=m_D$ for D$_2$ molecule). 
$\theta$ 
is the molecular orientation angle, 
where $\theta$=0 corresponds to the configuration of the molecule 
perpendicular 
to the surface 
while $\theta=\pi/2$ to the configuration parallel to the surface. 
$\hat{\vec{L}}$ is the associated angular momentum operator. 
$I(s)$ is the momentum inertia for the rotational motion given by 
\begin{equation}
I(s)=\mu r_0^2(1+f e^{\alpha s}), 
\label{mominertia}
\end{equation}
where $\mu=m/2$ and $f$ is a parameter characterizing the $s$ dependence 
of the interatomic distance $r$, $r_0$ being the interatomic 
distance for an isolate molecule. 
The same parameter $\alpha$ as in Eq. (\ref{mominertia}) 
appears also in the potential energy, 
$V(s,\theta)$, which is parametrised as 
\begin{eqnarray}
V(s,\theta)&=&\frac{E_a}{\cosh^2(\alpha s)}(1-\beta\cos\theta^2)
+V_1\cos^2\theta\cdot\frac{1}{2}(1+\tanh(\alpha s)), \\
&\equiv&V_0(s)+V_2(s)Y_{20}(\theta),
\label{Vmolecule}
\end{eqnarray}
with 
\begin{eqnarray}
V_0(s)&=&\frac{E_a}{\cosh^2(\alpha s)}\left(1-\frac{\beta}{3}\right)
+\frac{V_1}{6}(1+\tanh(\alpha s)), \\
V_2(s)&=&
-\frac{E_a}{\cosh^2(\alpha s)}\cdot \frac{2\beta}{3}\sqrt{\frac{4\pi}{5}}
+\frac{1}{3}\sqrt{\frac{4\pi}{5}}V_1(1+\tanh(\alpha s)). 
\end{eqnarray}

\begin{figure}[t]
\centerline{\includegraphics[width=11cm,clip]{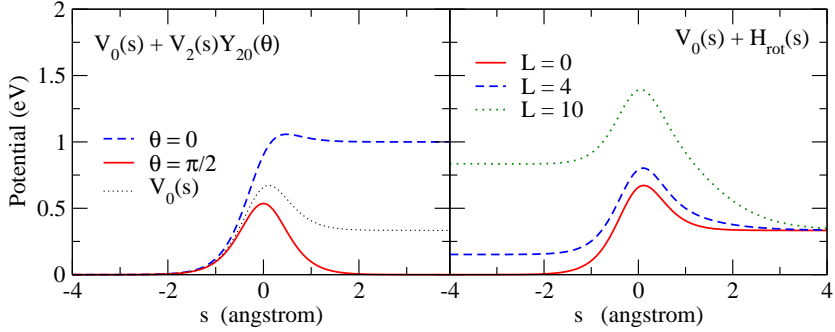}}
\caption{
The potential energy for the dissociative adsorption process of 
H$_2$ molecule on metal surface given by Eq. (\ref{Vmolecule}). 
The parameters are 
$E_a$=0.536 eV,  
$V_1$=1.0 eV, $\alpha$=1.5 \AA$^{-1}$, $\beta=0.25$, $r_0$=0.739 \AA, 
and $f=0.14$. 
The left panel shows the potential for $L=0$ 
as a function of the reaction path 
coordinate $s$ for $\theta=0$ (the dashed line) and $\theta=\pi/2$ (the solid 
line), where $\theta$ is 
the molecular orientation angle 
($\theta$=0 and $\theta=\pi/2$ correspond to the configurations with 
perpendicular and parallel to the metal surface, respectively) 
and $\hat{\vec{L}}$ is the associated angular momentum operator. 
The spherical part of the potential, $V_0(s)$, is also shown 
by the dotted line. 
The right panel shows the sum of spherical part of the potential, $V_0(s)$, 
and 
the rotational energy, $H_{\rm rot}(s)=L(L+1)\hbar^2/2I(s)$, for 
three different values of $L$. 
}
\end{figure}

The coupled-channels equations for the Hamiltonian (\ref{Hmolecule}) can 
be derived in the same manner as in Sec. 3. 
For scattering with the initial rotational angular momentum 
of molecules of $L_i$ and 
its $z$-component $M_i$, 
we expand the total wave function as 
\begin{equation}
\Psi_{L_iM_i}(s,\theta)=\sum_L\phi_{LL_i}(s)Y_{LM_i}(\theta).
\end{equation}
Notice that the Hamiltonian 
(\ref{Hmolecule}) conserve the value of $M_i$, as the coupling 
potential is proportional to $Y_{20}(\theta)$. 
The coupled-channels equations then read,
\begin{equation}
\left[-\frac{\hbar^2}{2M}\frac{d^2}{ds^2}+\frac{L(L+1)\hbar^2}{2I(s)}
+V_0(s)-E\right]\phi_{LL_i}(s)
+V_2(s)\sum_{L'}\langle Y_{LM_i}|Y_{20}|Y_{L'M_i}\rangle
\phi_{L'L_i}(s),
\label{ccmolecule}
\end{equation}
where the matrix element 
$\langle Y_{LM_i}|Y_{20}|Y_{L'M_i}\rangle$ is given by 
\begin{equation}
\langle Y_{LM_i}|Y_{20}|Y_{L'M_i}\rangle
=(-)^{M_i}\sqrt{\frac{5}{4\pi}}\sqrt{(2L+1)(2L'+1)}
\left(\begin{array}{ccc}
L&2&L'\\
0&0&0
\end{array}\right)
\left(\begin{array}{ccc}
L&2&L'\\
-M_i&0&M_i
\end{array}\right).
\label{couplingmolecule}
\end{equation}
Noticing that $I(s)\to \mu r_0^2$ for $s\to -\infty$ and 
$I(s)\to 0$ for $s\to \infty$, 
these coupled-channels equations are solved by imposing the boundary 
conditions of
\begin{eqnarray}
\phi_{LL_i}(s)&=& e^{ik_Ls}\delta_{L,L_i}
-\sqrt{\frac{k_{L_i}}{k_L}}\,R_{LL_i}
e^{-ik_Ls}~~~~~(s\to-\infty),    \\
&=&
\sqrt{\frac{k_{L_i}}{k}}\,T_{LL_i}
e^{iks}~~~~~~~~~~~~~~~~~~~~~~~~~(s\to\infty),
\end{eqnarray}
where 
$k_L=\sqrt{2M(E-\epsilon_L)/\hbar^2}$ with 
$\epsilon_L=L(L+1)\hbar^2/2\mu r_0^2$ and 
$k=\sqrt{2ME/\hbar^2}$. 
The adsorption probability for a given value of $L_i$ and $M_i$ is then 
obtained as 
\begin{equation}
P_{L_iM_i}=\sum_L|T_{LL_i}|^2.
\end{equation}
By making average over all possible $M_i$, the total adsorption probability 
for $L_i$ is given by 
\begin{equation}
P_{L_i}=\frac{1}{2L_i+1}\sum_{M_i}
P_{L_iM_i}.
\end{equation}

\begin{figure}[t]
\centerline{\includegraphics[width=11cm,clip]{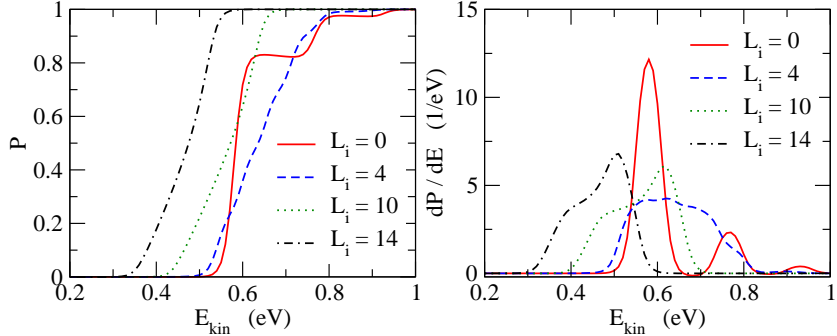}}
\caption{The results of the coupled-channels calculation for 
the dissociative adsorption process of H$_2$ molecules. 
The left panel shows the adsorption probability, $P$, while the 
right panels shows the barrier distribution defined as $dP/dE$ 
for several values of the initial angular momenta $L_i$ for the rotational 
state of the molecule as a function of the initial kinetic 
energy $E_{\rm kin}$. 
}
\end{figure}

Let us now solve the coupled-channels equations for H$_2$ molecules. 
The results are qualitatively the same also for D$_2$ molecules. 
Following Ref. \citen{DKO95}, 
we take $E_a$=0.536 eV,  
$V_1$=1.0 eV, $\alpha$=1.5 \AA$^{-1}$, $\beta=0.25$, and $r_0$=0.739 \AA. 
For the factor $f$ in Eq. (\ref{mominertia}), we take $f=0.14$\cite{D12}. 
The potential with these parameters are shown in Fig. 12. 
The left panel shows the potential 
energy $V(s,\theta)$ given by Eq. (\ref{Vmolecule}) for two different 
values of $\theta$. For comparison, the figure also shows the spherical 
part of the potential, $V_0(s)$. One can see that 
the barrier is lower for the configuration parallel to the metal surface 
(that is, $\theta=\pi/2$) as compared to the configuration perpendicular 
to the surface, $\theta=0$. The right panel, on the other hand, shows 
the sum of the spherical part of the potential, $V_0(s)$, and 
the rotational energy, $H_{\rm rot}(s)=L(L+1)\hbar^2/2I(s)$, for 
three different values of $L$. 
Because of the $s$ dependence of the rotational moment of inertia, $I(s)$, 
the barrier height for the molecules incident from $s=-\infty$, 
that is, the difference between the energy at $s=0$ and 
that at $s=-\infty$ decreases as a function of $L$. 

\begin{figure}[t]
\centerline{\includegraphics[width=11cm,clip]{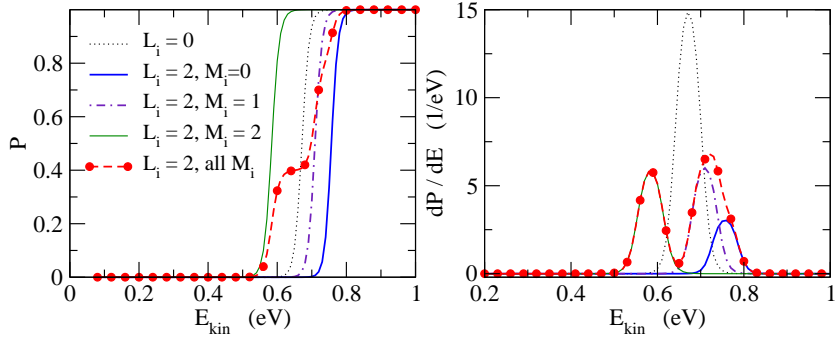}}
\caption{The results of 
the single-channel calculation, obtained by turning off all the coupling 
matrix elements in the coupled-channels equations except for the diagonal 
component. The left and the right panels show 
the adsorption probability, $P$, and 
the barrier distribution, $dP/dE$, respectively. 
The dotted lines denote the results when the initial rotational state 
is at $L_i=0$. The thin solid, the dot-dashed, and the thick solid lines 
are the results of $(L_i,M_i)=(2,0), (2,1)$, and (2,2), respectively. 
The dashed lines with the solid circles 
show the results for $L_i=2$ obtained by averaging all the 
$M_i$ components. For the barrier distributions shown in the right panel, 
the weight factors, 1/5 (for $M_i=0$) and 2/5 (for $M_i=1$ and 2), 
are multiplied. 
}
\end{figure}

The results of the coupled-channels calculations are shown in Fig. 13 for 
several values of initial rotational state, $L_i$, in which 
the adsorption probability is plotted as a function of the incident kinetic 
energy of the molecule, defined as $E=E_{\rm kin}+L_i(L_i+1)\hbar^2/2r_0^2$. 
As has been noted in 
Refs. \citen{DKO95} and \citen{DKO00}, these calculations well account 
for the non-monotonic behavior of the adsorption probability as a function 
of $L_i$. The right panel shows the corresponding 
barrier distribution, $dP/dE$, obtained with the point difference formula 
with the energy step of 0.03 eV. One can clearly see different structures 
for each $L_i$. For $L_i=0$, the barrier distribution has three prominent 
peaks. These peaks are smeared for $L_i=4$, and at the same time, the 
center of mass of the distribution is shifted towards high energy, leading 
to the decrease of adsorption probability. 
This is due to the fact that the result for $L_i=4$ is actually given 
by the average over contributions from 9 different $M_i$ values. 
In order to demonstrate this effect, Fig. 14 shows the results of the 
single-channel calculations for $L_i=2$ with three different values of 
$M_i$
%, that is, $M_i=0,1,$ and 2, 
and its average. 
For comparison, the figure also shows the single-channel calculation for 
$L_i=0$ (the dotted line). 
We define  
the single-channel calculation as the one which neglects all the coupling 
terms in the coupled-channels equations (\ref{ccmolecule}) 
except for the diagonal term, $L=L'$. 
Because of the properties of the spherical harmonics, the diagonal 
term of the coupling potential is attractive for $M_i=2$, while it is 
repulsive for $M_i=0$ and 1 (see Eq. (\ref{couplingmolecule})). 
The single peak in the barrier distribution 
for $L_i=0$ is then distributed to three peaks in the case of $L_i=4$, 
shifting the center of mass of the distribution slightly towards higher 
energy (notice that $-M_i$ gives the same contribution as $M_i$). 
With the off-diagonal components of the coupling potential, the 
distribution will be further smeared, as in the distribution for $L_i=4$ 
shown in Fig. 13. 
When the initial angular momentum is further increased, the barrier 
distribution starts moving towards lower energies, as seen in the figure for 
$L_i=10$ and 14, which enhances the adsorption probability as its 
consequence. This is mainly due to the fact that the barrier is lowered for 
a large value of rotational state, $L_i$, 
as has been shown in Fig. 12. 

The barrier distribution representation of the tunneling probability 
provides a useful means to 
understand the 
underlying dynamics of dissociative adsorption process, as the shape of 
the distribution strongly reflects the molecular intrinsic motions. 
This would 
be even so, particularly when 
the rotational and the vibrational degrees are taken into account 
simultaneously\cite{DKO96,MKD99}. 
It will be an interesting future study to investigate how the barrier 
distribution behaves in 
the presence of the rotational excitation together with the vibrational 
excitation. 

\section{Summary and outlook}

Recent developments in experimental techniques have enabled 
high precision measurements of heavy-ion fusion cross sections. 
Such high precision experimental data have elucidated 
the mechanism of subbarrier fusion reactions 
in terms of quantum tunneling 
of systems with many degrees of freedom. 
In particular, 
the effects of 
the coupling of the relative motion 
between the target and projectile 
nuclei to 
their intrinsic excitations 
have been transparently clarified through the barrier distribution 
representation of fusion cross sections. 

The channel coupling effects can be taken into account most naturally with 
the coupled-channels method. 
When the excitation energy of an intrinsic motion to which 
the relative motion couples is zero, 
the concept of barrier distribution holds exactly. 
In this case, quantum tunneling takes place much faster than the intrinsic 
motion. The effects of the couplings can then be 
expressed in terms of the distribution of potential barriers, and 
the fusion cross sections are given as a weighted sum of fusion 
cross sections for the distributed barriers. 
The underlying structure of the barrier distribution 
can be most clearly investigated when the 
first derivative of barrier penetrability, $dP/dE$, is plotted as a 
function of energy. In heavy-ion fusion reactions, 
this quantity corresponds to the second derivative of $E\sigma_{\rm fus}$, 
which is referred to as fusion barrier distribution, $D_{\rm fus}$. 
The fusion barrier distribution has been extracted for many systems 
through the high precision experimental data of fusion 
cross sections, $\sigma_{\rm fus}$. 

Even when the excitation energy of the intrinsic motion is not zero, 
the concept of fusion barrier distribution can be approximately generalized, 
using the eigen-channel representation of nuclear $S$-matrix, 
defined as the eigen-states of $S^\dagger S$. 
We have demonstrated that the barrier distribution shows 
a transition from the sudden to the adiabatic tunneling limits 
in a natural way as the excitation energy increases, 
where the potential is simply renormalized 
in the latter limit without affecting 
the shape of barrier distribution 
({\it i.e.,} the adiabatic barrier renormalization). 

The barrier distribution representation is applicable also to 
other multi-channel quantum tunneling problems. A good example 
is the dissociative adsorption phenomenon in surface science. 
The rotational and vibrational excitations of diatomic molecules play 
an important role in the adsorption process. These effects can be 
described by the coupled-channels approach, and the barrier distribution 
can be defined as in heavy-ion subbarrier fusion reactions. 
The results of coupled-channels calculations have indicated that 
the barrier distribution representation provides a useful means 
in clarifying the underlying mechanism in the dynamics of surface 
interaction of molecules. 

Although our understanding of subbarrier fusion reactions has 
considerably increased in the past decades, 
there are still many open problems in heavy-ion subbarrier fusion 
reactions. 
For example, 
it has not yet been understood completely how 
the hindrance of fusion cross sections 
with respect to the standard coupled-channels calculation 
takes place at deep subbarrier 
energies.  
A promising mechanism of the hindrance is that many non-collective 
channels are activated after the target and the projectile nuclei 
overlap with each other, and the relative energy is irreversibly 
dissipated to the intrinsic motions. This would occur only at deep 
subbarrier energies, in which the inner turning point of the potential 
barrier is located inside the touching radius of the two nuclei. 
This phenomenon may thus be 
a good example of dissipative quantum tunneling, which has been 
extensively discussed in many fields of physics and chemistry. 
A unique feature in nuclear physics is that the dissipative nature 
of the couplings gradually sets in, in a sense that the coupling is 
reversible before the touching and it gradually reveals the irreversible 
character as the overlap of the colliding nuclei increases. 
In order to gain a deep insight of this problem, it might be helpful 
to revisit heavy-ion deep inelastic collisions (DIC) from a more quantum 
mechanical point of view. 
This is important also 
in connection with the synthesis of superheavy elements 
by heavy ion collisions with large mass numbers, for which 
fusion cross 
is strongly hindered at energies near the bare Coulomb barrier.

Other important issues not covered in this paper include 
fusion of halo nuclei and the role of multi-nucleon transfer. 
For the former, there have been many debates concerning how 
the breakup process affects subbarrier 
fusion \cite{HPCD92, TKS93,DV94,HVDL00,DT02,IYNU06,GCLH11}. However 
the interplay between fusion and breakup involve many 
complex processes \cite{CGDH06} and the role of breakup in fusion has not 
yet been understood completely. Moreover, particle transfer processes 
also affect both fusion and breakup in a non-trivial way, as has been 
found recently in $^{6,7}$Li + $^{208}$Pb reactions\cite{LDH11} (see 
Ref. \citen{DHB99} for a review on 
subbarrier fusion of weakly bound stable nuclei, 
$^{6,7}$Li and $^9$Be). 
A theoretical calculation has to take into account the fusion, 
transfer, and breakup processes simultaneously in a consistent manner. 
It remains a challenging problem to carry out such calculations, although 
the time-dependent wave packet 
approach \cite{IYNU06} has been performed 
with a limited partition for the 
transfer channels.
From experimental side, fusion cross sections for many 
neutron-rich nuclei do not appear to 
show any particular enhancement or hindrance \cite{R04,S04,L09,LGK12}, but 
recent experimental data for $^{12,13,14,15}$C+$^{232}$Th reactions 
have shown that the fusion cross sections are enhanced for the $^{15}$C 
projectile as compared to those with the other C isotopes \cite{A11}. 
Again, several sort of transfer channels would have to be considered to 
understand the differences in the behavior of fusion cross 
sections \cite{LDH11,L11,RRL10,KLS11}. 
In particular, the multi-nucleon transfer process may play an 
important role in fusion of neutron-rich nuclei. 
Although there have been a few attempts to treat 
the multi-neutron transfer process in subbarrier fusion reactions
\cite{RTN92,EJR98,R01,PW00,Z03}, it has still been a challenging problem 
to include in a full quantum mechanical manner 
the multi-nucleon transfer processes consistently 
with inelastic channels by taking into account also the 
final $Q$-value distribution of transfer. 

A much more challenging problem is to describe heavy-ion fusion 
reactions, thus many-particle tunneling\cite{B61}, 
from fully quantum many-body perspectives, starting from nucleon 
degrees of freedom. 
The time-dependent Hartree-Fock (TDHF) theory has been 
widely employed to microscopically describe nuclear dynamics 
\cite{RS80,SLZ10}. It has been well known, however, that the TDHF method has 
a serious drawback that it cannot describe a many-particle tunneling 
phenomenon. In order to cure this problem, 
Bonasera and Kondratyev have introduced 
the imaginary time propagation \cite{BK94,KBI00}. 
In this connection, we wish to mention that 
an alternative imaginary time approach, called 
the mean field tunneling theory, for quantum 
tunneling of systems with many degrees of freedom has been 
developed in Ref. \citen{KTAB03}. 
The mean field tunneling theory
is a reformulation of the 
dynamical norm method for quantum tunneling \cite{THA95,HB04}, 
which 
evaluates the non-adiabatic effect on the tunneling rate 
through the change of the norm of the wave function 
for the intrinsic space 
during the evolution along 
the imaginary time axis.  
The mean field tunneling theory 
has been applied to quantum mechanically 
discuss electron screening effects in low energy 
nuclear reactions\cite{KTAB03}, while the dynamical norm 
method has been used to 
discuss the effects of nuclear oscillation on fission \cite{THA95}. 
It would be an interesting challenge to develop 
a fully microscopic version of these methods and apply them 
to heavy-ion fusion reactions.
More recently, Umar {\it et al.} have used the density constrained 
TDHF (DC-TDHF) method to analyze heavy-ion 
fusion reactions\cite{UO07,OUMR10,KUO12}. 
Even though these microscopic approaches seem promising, 
these are based on certain assumptions, such as a 
local collective potential
with single-channel. It is thus not yet clear whether they are 
applicable to many-particle tunneling problems in general, such as 
two-proton radioactivity \cite{Gri01,Gri09,PKGR12,BP08,G09,MOHS12} 
and alpha decays\cite{D10,SKT70,TonAri79,VarLio94,BN12}. 
It would be an ultimate goal to develop a general microscopic theory which 
can describe several tunneling phenomena simultaneously, 
not only in nuclear physics but 
also in other fields of physics and chemistry. Such theory would 
naturally provide 
a way to describe the role of irreversibility (that is, the energy 
and angular momentum dissipations) as well as the density evolution after 
the touching in subbarrier fusion reactions without any assumption for the 
adiabaticity of the fusion process. 

\section*{Acknowledgements}
We thank D.M. Brink, A.B. Balantekin, N. Rowley, A. Vitturi, 
M. Dasgupta, D.J. Hinde, T. Ichikawa, 
M.S. Hussein, L.F. Canto, 
C. Beck, L. Corradi, A. Diaz-Torres, 
P.R.S. Gomes, S. Kuyucak, J.F. Liang, 
C.J. Lin, G. Montagnoli, A. Navin, 
G. Pollarolo, F. Scarlassara, A.M. Stefanini, and H.Q. Zhang 
for collaborations and many
useful discussions.
K.H. also thanks Y. Miura, T. Ichikawa, W.A. Di\~no, and S. Suto 
for useful discussions on dissociative adsorption in surface physics. 
This work was supported
by the Japanese
Ministry of Education, Culture, Sports, Science and Technology
by Grant-in-Aid for Scientific Research under
program no. (C) 22540262.

\appendix
\section{Relation between 
surface diffuseness and barrier parameters}

In this Appendix, we discuss the relation between the surface 
diffuseness parameter $a$ 
in a nuclear potential and the parameters which characterise 
the Coulomb barrer, that is, the curvature, the barrier height, and the 
barrier position. 
With such relation, one can estimate the value of $a$ from empirical 
barrier parameters. 

For a given nuclear potential $V_N(r)$, the barrier position $R_b$ is 
obtained from the condition that the first derivative of 
the total potential is zero at $r=R_b$, 
\begin{equation}
\left.\frac{d}{dr}V(r)\right|_{r=R_b} = \left[
\frac{dV_N(r)}{dr}-\frac{Z_PZ_Te^2}{r^2}\right]_{r=R_b}=0.
\label{barrier}
\end{equation}
The barrier height $V_b$ and the curvature $\Omega$ are then 
evaluated as 
\begin{eqnarray}
V_b&=&V_N(R_b) + \frac{Z_PZ_Te^2}{R_b}, 
\label{barrier2}
\\
\Omega &=& \sqrt{-\frac{V_N^{''}(R_b)+2Z_PZ_Te^2/R_b^3}{\mu}},
\label{barrier3}
\end{eqnarray}
where $V_N^{''}(r)$ is the second derivative of the nuclear 
potential with respect to $r$. 

\subsection{Exponential potential}

We first consider an exponential potential given by 
\begin{equation}
V_N(r)=V_0e^{-r/a}.
\end{equation}
From Eq. (\ref{barrier}), the depth of the nuclear potential, $V_0$, is 
related to the charge product $Z_PZ_T$ as 
\begin{equation}
-\frac{V_0}{a}e^{-R_b/a}-\frac{Z_PZ_Te^2}{R_b^2}=0. 
\end{equation}
From this equation,  
the barrier height and 
the curvature read 
\begin{eqnarray}
V_b&=& \frac{Z_PZ_Te^2}{R_b}\left(1-\frac{a}{R_b}\right), \\
\Omega^2&=&\frac{Z_PZ_Te^2}{\mu R_b^2}\left(\frac{1}{a}-\frac{2}{R_b}
\right),
\end{eqnarray}
respectively. 

\subsection{Woods-Saxon potential}

We next consider a Woods-Saxon potential given by 
\begin{equation}
V_N(r)=-\frac{V_0}{1+e^{(r-R_0)/a}}.
\end{equation}
Combining Eqs. (\ref{barrier}),(\ref{barrier2}), and (\ref{barrier3}), 
one finds that the surface diffuseness 
parameter $a$ is expressed in terms of $R_b, V_b$ and $\Omega$ as 
\begin{equation}
a=\frac{R_b}{-\frac{\mu\Omega^2R_b^3}{Z_PZ_Te^2}
-2 + \frac{2Z_PZ_Te^2}{Z_PZ_Te^2-R_bV_b}}.
\end{equation}
Once the surface diffuseness parameter is thus 
evaluated, the other two 
parameters in the nuclear potential can be obtained as 
\begin{eqnarray}
1+e^{-x}&=& \frac{1}{a}\frac{R_b^2}{Z_PZ_Te^2}
\left(\frac{Z_PZ_Te^2}{R_b}-V_b\right), \\
V_0&=& a e^{-x}(1+e^x)^2\frac{Z_PZ_Te^2}{R_b^2},
\end{eqnarray}
where $x$ is defined as $(R_b-R_0)/a$.

\section{Parabolic approximation and the Wong formula}

If the Coulomb barrier 
is approximated by a parabola, 
\begin{equation}
V(r)\sim V_b -\frac{1}{2}\mu\Omega^2(r-R_b)^2,
\end{equation}
the corresponding penetrability can be 
evaluated analytically as 
\begin{equation}
P(E)=\frac{1}{1+\exp\left[\frac{2\pi}{\hbar\Omega}(V_b-E)\right]}. 
\end{equation}
Using the parabolic approximation, Wong has derived an analytic 
expression for fusion cross sections \cite{W73}. 
He assumed that (i) the curvature of the Coulomb barrier, $\hbar\Omega$, 
is 
independent of the angular momentum $l$, and (ii) 
the position of the Coulomb barrier, $R_b$, 
is also independent of $l$, and 
the dependence 
of the penetrability on the 
angular momentum can be well approximated 
by shifting the incident energy as 
\begin{equation}
P_l(E)=P_{l=0}\left(E-\frac{l(l+1)\hbar^2}{2\mu R_b^2}\right). 
\label{ldepP}
\end{equation}
If many partial waves contribute to fusion cross section, 
the sum in Eq. (\ref{fuscross2}) may be replaced by an integral, 
\begin{equation}
\sigma_{\rm fus}(E)=\frac{\pi}{k^2}\int_0^{\infty}dl~(2l+1)P_l(E).
\label{wong0}
\end{equation}
Changing the variable from $l$ to $l(l+1)$, the integral 
can be explicitly carried out, leading to the Wong formula \cite{W73}
\begin{equation}
\sigma_{\rm fus}(E)=\frac{\hbar\Omega}{2E}R_b^2 \ln\left[1+\exp\left(
\frac{2\pi}{\hbar\Omega}(E-V_b)\right)\right]. 
\label{wong}
\end{equation}
At energies well above the Coulomb barrier, this formula reduces to 
the classical expression of fusion cross section given by 
Eq. (\ref{classical}). 
%
%\begin{equation}
%\sigma(E)=\pi R_b^2\left(1-\frac{V_b}{E}\right) ~~~~~~(E \gg V_b).
%\label{classical}
%\end{equation}
%

\begin{figure}[t]
\centerline{\includegraphics[width=11cm,clip]{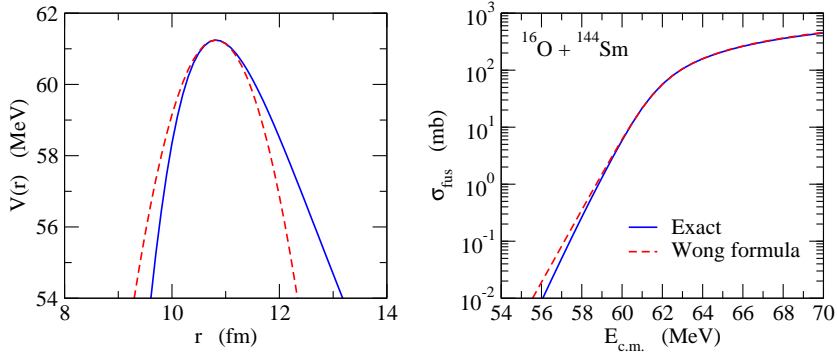}}
\caption{(The left panel) The Coulomb barrier for the 
$^{16}$O+$^{144}$Sm system shown in Fig. 1 (the solid line) and 
its parabolic approximation (the dashed line). 
(The right panel) Comparison of the 
corresponding 
fusion cross sections obtained by 
numerically solving the Schr\"odinger equation 
without resorting to the parabolic approximation 
(the solid line) and 
those obtained with the Wong formula, Eq. (\ref{wong}). 
}
\label{fig:app-b}
\end{figure}

The left panel of Fig. \ref{fig:app-b} shows the parabolic approximation 
to the Coulomb barrier for the 
$^{16}$O + $^{144}$Sm system shown in Fig. 1. 
Because of the long ranged Coulomb interaction, the Coulomb barrier is 
asymmetric and the parabolic potential has a smaller width compared 
with the realistic potential. Nevertheless, the Wong formula for 
fusion cross sections, Eq. (\ref{wong}), works well except at energies 
well below the barrier, where the parabolic approximation breaks down 
(see the right panel of Fig. \ref{fig:app-b}). 

\begin{figure}[tb]
\centerline{\includegraphics[width=7cm,clip]{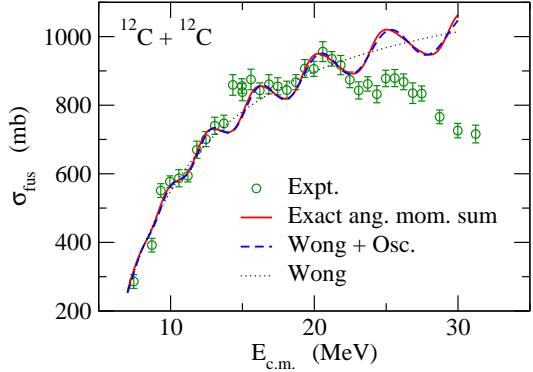}}
\caption{The fusion excitation function for the $^{12}$C+$^{12}$C 
system. The solid line is obtained by carrying out exactly 
the angular momentum 
sum (with the symmetrization factor) 
in Eq. (\ref{fuscross2}) with a 
parabolic potential with 
$V_b=5.6$ MeV, $R_b$=6.3 fm, and $\hbar\Omega=3$ MeV. 
The barrier position and the curvature are assumed to be independent of 
the angular momentum $l$. 
The dotted line is obtained with 
the Wong formula, Eq. (\ref{wong}), while the dashed 
line is obtained as 
a sum of the Wong formula and the oscillatory 
cross sections given by Eq. (\ref{osc2}). 
The experimental data are taken from Ref. \citen{S76}.}
\label{fig:app-b2}
\end{figure}

Even though the Wong formula appears to work well for the single-channel 
potential model, 
one can still discuss the corrections to it. 
The first correction is with respect to 
the integral in Eq. (\ref{wong0}). 
To discuss the correction, we first notice that 
replacing the sum in Eq. (\ref{fuscross2}) with the integral in 
Eq. (\ref{wong0}) is equivalent to taking only the leading term ($m=0$) of 
the exact Poisson sum formula, 
\begin{equation}
\sigma_{\rm fus}(E)=\frac{\pi}{k^2}\sum_l(2l+1)P_l(E)
=\frac{2\pi}{k^2}\sum_{m=-\infty}^\infty\int^\infty_0\lambda 
P(E;\lambda)e^{2\pi mi\lambda}d\lambda,
\end{equation}
where $P(E;\lambda)$ is any smooth function of $\lambda$ satisfying 
$P(E,l+1/2)=P_l(E)$ \cite{Brink85}.
Poffe, Rowley, and Lindsay have evaluated the contribution of 
the next most important terms, $m=\pm 1$\cite{PRL83}. These terms lead 
to an oscillatory contribution to the fusion cross sections, 
\begin{equation}
\sigma_{\rm fus}(E)=\sigma_W(E)+\sigma_{\rm osc}(E),
\end{equation}
where $\sigma_W(E)$ is given by Eq. (\ref{wong}), while the oscillatory 
part $\sigma_{\rm osc}(E)$ is given by
\begin{equation}
\sigma_{\rm osc}(E)=4\pi\mu R_b^2\frac{\hbar\Omega}{k^2}\exp\left(-
\frac{\pi\mu R_b^2\hbar\Omega}{l_g+\frac{1}{2}}\right)\sin(2\pi l_g). 
\end{equation}
Here, $l_g$ is the grazing angular momentum satisfying 
\begin{equation}
E=V(r)+\frac{l_g(l_g+1)\hbar^2}{2\mu R_b^2}.
\end{equation}
For heavy systems, 
the oscillatory part of fusion cross sections, $\sigma_{\rm osc}$, 
is usually much smaller than the leading term, $\sigma_W$. 
However, for light symmetric systems such as $^{12}$C+$^{12}$C, 
the oscillatory part becomes significant\cite{PRL83,KKR83,E12,S76,T78}. 
For a system of identical spin-zero bosons, the factor $(1+(-1)^l)$ has 
to be multiplied in the sum in Eq. (\ref{fuscross2}) due to the 
symmetrization effect, making the contributions of 
all the odd partial waves vanish. 
In this case, the leading term of the fusion cross section is still 
given by the Wong formula, 
Eq. (\ref{wong}), while the oscillatory part becomes \cite{PRL83}
\begin{equation}
\sigma_{\rm osc}(E)=4\pi\mu R_b^2\frac{\hbar\Omega}{k^2}\exp\left(-
\frac{\pi\mu R_b^2\hbar\Omega}{2l_g+1}\right)\sin(\pi l_g). 
\label{osc2}
\end{equation}
Fig. \ref{fig:app-b2} shows the fusion cross sections 
for the $^{12}$C+$^{12}$C reaction obtained with a 
parabolic potential with 
$V_b=5.6$ MeV, $R_b$=6.3 fm, and $\hbar\Omega=3$ MeV. 
The solid line shows the result of the exact summation of partial wave 
contributions with Eq. (\ref{ldepP}), 
while the dashed line shows a sum of Eqs. (\ref{wong}) and 
(\ref{osc2}). A separate contribution from the Wong formula, Eq. (\ref{wong}), 
is also shown by the dotted line. It is seen that the oscillation of fusion 
cross sections can be well reproduced with the formula given by 
Eq. (\ref{osc2}). 

The second correction to the Wong formula is the angular momentum 
dependence of the barrier radius\cite{BDK96}. 
Up to the first order of 
$\hbar^2/\mu^2\Omega^2 R_b^4$, 
Balantekin {\it et al.} have shown 
that the barrier radius 
for the $l$-th partial wave $R_{bl}$ is given by 
\begin{equation}
R_{bl}=R_b-\frac{l(l+1)\hbar^2}{\mu^2\Omega^2 R_b^3}.
\end{equation}
This equation indicates that the barrier position decreases 
as the angular momentum $l$ increases. 
At energies well above the barrier, the classical fusion cross 
sections are then modified to \cite{BDK96} 
\begin{equation}
\sigma_{\rm fus}(E)=\pi R_b^2\left(1-\frac{V_b}{E}\right)
-\frac{2\pi}{\mu\Omega^2E}(E-V_b)^2
 ~~~~~~(E \gg V_b).
\label{classical2}
\end{equation}
Comparison between Eqs. (\ref{classical}) and (\ref{classical2}) 
shows that 
the Wong formula slightly overestimates fusion cross sections 
at energies well above the Coulomb barrier. 

\section{Multiphonon coupling}

In this Appendix, we show that the dimension of coupled-channels equations 
can be reduced for vibrational couplings 
by introducing effective multi-phonon channels. 
Suppose that we have two modes of vibrational excitations ({\it e.g.,} 
a quadrupole and an octupole modes), 
and consider the excitation operator 
\begin{equation}
\hat{O}=\beta_1(a_1^\dagger+a_1)+\beta_2(a_2^\dagger+a_2), 
\end{equation}
and the phonon Hamiltonian
\begin{equation}
H_0=\hbar\omega_1a_1^\dagger a_1+\hbar\omega_2a_2^\dagger a_2,
\end{equation}
where $a_1^\dagger$ and $a_2^\dagger$ are the phonon creation operator 
for the first and the second modes, respectively. 
$\beta_i~(i=1,2)$ are the coupling constants, while 
$\hbar\omega_i~(i=1,2)$ are the phonon excitation energy for each mode. 
We have shifted the phonon energies so that the ground state 
is at zero energy. 

If we truncate the phonon space up to 
the one phonon states, 
we have three basis states, 
$|00\rangle, |10\rangle$, and $|01\rangle$, 
where 
the state $|n_1n_2\rangle$ corresponds to the 
product state of $n_1$ phonon state for the first mode 
and $n_2$ phonon state for the second mode. Here we have included the 
states with $n_1+n_2\leq 1$.  
The matrix elements of the operator $H_0+\hat{O}$ 
with these basis states read, 
\begin{equation}
H_0+\hat{O}=\left(
\begin{array}{ccc}
0&\beta_1&\beta_2\\
\beta_1&\hbar\omega_1&0\\
\beta_2& 0 & \hbar\omega_2\\
\end{array}
\right),
\end{equation}
It is easy to see that the ground state $|00\rangle$ couples only to a 
particular combination of $|10\rangle$ and $|01\rangle$\cite{KRNR93}, 
\begin{equation}
|\tilde{1}\rangle=\frac{1}{\sqrt{\beta_1^2+\beta_2^2}}\,(\beta_1|10\rangle 
+\beta_2|01\rangle),
\end{equation}
with
\begin{equation}
\hat{O}|00\rangle=\sqrt{\beta_1^2+\beta_2^2}\,|\tilde{1}\rangle. 
\end{equation}
The other combination of $|10\rangle$ and $|01\rangle$, 
$\beta_2|10\rangle - \beta_1|01\rangle$, couples neither to $|00\rangle$ 
nor $|\tilde{1}\rangle$, 
and this can be removed from the coupled-channels 
calculation if the excitation 
energies of the two modes are the same, $\hbar\omega_1
=\hbar\omega_2\equiv \hbar\omega$. 
In this case, the dimension of the coupled-channels equations 
can be reduced to two with a modified strength as\cite{KRNR93},
\begin{equation}
H_0+\hat{O}=\left(
\begin{array}{cc}
0&\bar{\beta}\\
\bar{\beta}&\hbar\omega
\end{array}
\right),
\end{equation}
where $\bar{\beta}$ is defined by 
$\bar{\beta}=\sqrt{\beta_1^2+\beta_2^2}$.
One can easily generalize this scheme to higher members of phonon states. 
The resultant matrix is equivalent to that for a single phonon mode 
with the effective strength $\bar{\beta}$. 
For instance, when the phonon space is truncated at the two-phonon states, 
the coupling matrix reads
\begin{equation}
H_0+\hat{O}=\left(
\begin{array}{ccc}
0&\bar{\beta}&0\\
\bar{\beta}&\hbar\omega&\sqrt{2}\bar{\beta} \\
0&\sqrt{2}\bar{\beta}&2\hbar\omega
\end{array}
\right),
\end{equation}
where the effective two phonon state is defined as 
\begin{equation}
|\tilde{2}\rangle=\frac{1}{\beta_1^2+\beta_2^2}\left(\beta_1^2|20\rangle 
+\sqrt{2}\beta_1\beta_2|11\rangle+\beta_2^2|02\rangle\right). 
\end{equation}
%

%\appendix
%\section{First Appendix} %Empty argument \section{} yields `Appendix'. 
%
%\section{Second Appendix}

\end{document}